\documentclass[journal,transmag]{IEEEtran}
\usepackage{graphicx}
\usepackage{float}
\usepackage{amsmath}
\usepackage{amsfonts}
\usepackage{multirow}
\usepackage{subfigure}
\usepackage{caption}\usepackage{subcaption}
\usepackage{color}
\usepackage{comment}
\newcommand {\Define} {\stackrel {\Delta} {=}  }

\begin{document}
%

\title{OTFS - Predictability in the Delay-Doppler Domain and its Value to Communication and Radar Sensing}


\author{\IEEEauthorblockN{Saif Khan Mohammed\IEEEauthorrefmark{1},
Ronny Hadani\IEEEauthorrefmark{2},
Ananthanarayanan Chockalingam\IEEEauthorrefmark{3}, and 
Robert Calderbank\IEEEauthorrefmark{4},~\IEEEmembership{Fellow,~IEEE}}
\IEEEauthorblockA{\IEEEauthorrefmark{1}Department of Electrical Engineering, Indian Institute of Technology Delhi, India}
\IEEEauthorblockA{\IEEEauthorrefmark{2}Department of Mathematics, University of Texas at Austin, USA}
\IEEEauthorblockA{\IEEEauthorrefmark{3} Department of Electrical and Communication Engineering, Indian Institute of Science Bangalore, India}
\IEEEauthorblockA{\IEEEauthorrefmark{4}Department of Electrical and Computer Engineering, Duke University, USA}
\thanks{This article has been accepted for publication in IEEE BITS the Information Theory Magazine. DOI: 10.1109/MBITS.2023.3319595}
\thanks{$\copyright$ 2023 IEEE. Personal use of this material is permitted. Permission
from IEEE must be obtained for all other uses, in any current or future
media, including reprinting/republishing this material for advertising or
promotional purposes, creating new collective works, for resale or
redistribution to servers or lists, or reuse of any copyrighted
component of this work in other works.}
}

\IEEEtitleabstractindextext{
\begin{abstract}
In our first paper \cite{PartIpaper} we explained why the Zak-OTFS input-output (I/O) relation is predictable
and non-fading when the delay and Doppler periods are greater than the effective channel delay and Doppler spreads, a condition which we refer to as the crystallization condition. We argued that a communication system should operate within the crystalline regime.

{It is well known that it is possible to identify a linear time varying (LTV) channel if and only if it is under-spread. The crystallization condition is more restrictive than the under-spread condition, so identification is always possible. In the crystalline regime, we show that Zak-OTFS pilot sequences minimize the complexity of identifying the effective DD domain channel filter. We demonstrate that the filter taps can simply be read off from the response to a single Zak-OTFS pilot.}
In general, we provide an explicit formula for reconstructing the Zak-OTFS I/O relation from a finite number of received pilot symbols in the delay-Doppler (DD) domain. This reconstruction formula makes it possible to study predictability of the Zak-OTFS I/O relation for a sampled system that operates under finite duration and bandwidth constraints. We analyze reconstruction accuracy for different choices of the delay and Doppler periods, and of the pulse shaping filter. Reconstruction accuracy is high when the crystallization condition is satisfied, implying that it is possible to learn directly the I/O relation without needing to estimate the underlying channel. This opens up the possibility of a model-free mode of operation, which is especially useful when a traditional model-dependent mode of operation (reliant on estimation of the underlying physical channel) is out of reach (for example, when the channel comprises of unresolvable paths, or exhibits a continuous delay-Doppler profile such as in presence of acceleration).
Our study clarifies the fundamental origins of predictability by revealing how non-predictability
appears as a consequence of aliasing in the DD domain. This perspective leads to a canonical decomposition of the effective DD channel as a sum of predictable and non-predictable components, which we refer to as the crystalline decomposition. Vanishing of the non-predictable component of the channel is equivalent to satisfying the crystallization condition.

Finally, we measure the benefits of predictability in terms of bit error rate (BER) performance. We consider two cases. In the first, we measure performance given perfect knowledge of the I/O relation. We show that performance is optimal when the crystallization condition holds, that performance approaches that of {Time Domain Modulation (TDM)} when the Doppler period vanishes, and approaches that of {Frequency Domain Modulation (FDM)} when the delay period vanishes. In the second, we measure performance given imperfect knowledge of the I/O relation, as is the case when it is not possible to learn the underlying channel. We show that model-free operation is successful when the crystallization condition holds, and that performance is only slightly worse than performance given perfect knowledge of the I/O relation.

We also compare the performance of Zak-OTFS with that of a well-studied conventional multi-carrier approximation to Zak-OTFS, which we refer to as MC-OTFS. We show that the I/O relation of MC-OTFS is predictable to a lesser degree than that of Zak-OTFS, and as
a result the performance of MC-OTFS is inferior as the Doppler spread increases.
\end{abstract}

}

\maketitle

\IEEEdisplaynontitleabstractindextext

\IEEEpeerreviewmaketitle

%
%
%
%


\section{Introduction}
\label{secIntro}
6G presents an opportunity to reflect on the fundamentals of wireless communication, as it becomes more and more difficult to estimate channels, and we encounter Doppler spreads measured in KHz (e.g., 1.3 KHz Doppler at 28 GHz carrier and 50 km/hr speed, and 2.3 KHz Doppler at 5 GHz carrier and 500 km/hr speed) \cite{6Gpaper}. It is even an opportunity to question the standard model-dependent approach to wireless communication that requires channel estimation.

It is common knowledge that a time-domain (TD) pulse is an ideal waveform for pure delay channels
as it is possible to separate reflections according to their range, and, similarly, a frequency domain (FD) pulse is an ideal waveform for pure Doppler channels as it is possible to separate reflections according to their velocity. In part II of this tutorial paper, we explore the proposition that
a pulse in the delay-Doppler (DD) domain is an ideal waveform for doubly spread channels comprising of reflections of various ranges and velocities. In Part I \cite{PartIpaper}, we explained that a pulse
in the DD domain is a quasi-periodic localized function, and that when viewed in the time domain,
is realized as a pulse train modulated by a tone, (hence the name \emph{pulsone}).


In Part I, we described a modulation scheme referred to as Zak-OTFS, which uses the
inverse Zak transform \cite{Zak67, Janssen88} to convert information symbols mounted on DD pulses to the time domain for transmission. 
{The Zak transform converts a TD signal to its DD realization which is parameterized by the delay period $\tau_p$ and the Doppler period $\nu_p = 1/\tau_p$. The DD realization is quasi-periodic, with period $\tau_p$ along the delay axis and $\nu_p$ along the Doppler axis.}

{The Zak transform appears in the signal processing literature as method of analyzing signals (see \cite{Janssen88}), and not as a modulation technique. This was also true of the Fourier transform, until the introduction of OFDM in the 1970s, and our treatment of Zak - OTFS is inspired by this historical development.}

In Part I,  we emphasized that the Zak-OTFS input-output (I/O) relation is predictable and non-fading
when the delay and Doppler periods are greater than the effective channel delay and Doppler spreads,
a condition we call the \emph{crystallization condition}. {We argued that to achieve robust performance, a Zak-OTFS based communication system should operate within the \emph{crystalline regime}, i.e., the segment of the rectangular hyperbola $\tau_p \, \nu_p = 1$ where $(\tau_p, \nu_p)$ satisfy the crystallization condition.} {The Zak-OTFS I/O relation captures the interaction of the effective channel with the Zak-OTFS modulation. When the crystallization condition holds, the I/O relation can be determined by measuring the effect of the underspread channel on a single Zak-OTFS pulsone.}

In Part I, we described the predictability and non-fading attributes of Zak-OTFS in the context of continuous time and infinite bandwidth. Here in Part II, we study predictability in the context of a sampled communication system with finite duration and bandwidth constraints, and we present a discrete DD domain system model that enables bit error rate (BER) performance evaluation through simulation. We measure predictability through an explicit formula for reconstructing the I/O relation from a finite number of received pilot samples in the DD domain. The reconstruction accuracy depends on the choice of the delay-Doppler periods and the pulse shaping filters, and accuracy is high when the crystallization condition is satisfied. In the crystalline regime, it is possible to learn the I/O relation without needing to estimate the underlying channel. This opens up the possibility of model-free operation, which can enable communication when traditional model-dependent modes requiring channel estimation are out of reach (for example when the channel comprises of non-resolvable
paths, or admits a continuous delay-Doppler profile, as in the presence of acceleration). We now highlight our main contributions.

\textbf{Origins of non-predictability}:
We show that non-predictability and fading result from aliasing in the DD domain, and we describe how aliasing occurs when the channel delay spread is greater than the delay period, or the channel Doppler spread is greater than the Doppler period. Fundamental understanding of aliasing leads to the \emph{crystalline decomposition}, which is a canonical decomposition of the effective channel response filter into a predictable component and a non-predictable component. The crystallization condition
holds if and only if the non-predictable component vanishes.

\textbf{Benefits of predictability}:
Given the I/O response at one DD domain point in a frame, it is possible to predict the I/O response at all other points in the frame. Predictability implies that the received power profile is flat (no fading), that we have engineered a two-dimensional Gaussian channel. We illustrate the practical benefits of predictability by evaluating BER performance as a function of the received SNR.
We first suppose that channel estimation is perfect in order to understand the
impact of fading. We show that performance is superior in the crystalline regime, that performance approaches that of {Time Domain Modulation (TDM)} as the Doppler period shrinks, and that performance approaches that of {Frequency Domain Modulation (FDM)} as the delay period shrinks. {Note that information is carried by TD pulses in TDM and by FD pulses in FDM.} We then evaluate performance when we do not have the fine delay-Doppler resolution necessary for accurate channel estimation, and as a consequence, model-dependent approaches fail.
We show that in the crystalline regime, model-free operation is successful,
and that performance is only slightly worse than performance with perfect knowledge of the effective channel. We also describe how better transmit and receive filters serve to extend the range of reliable operation.

\textbf{Optimality of Zak-OTFS}:
Over the past few years 
several variants of OTFS have been reported in literature \cite{TThaj2022}. A multicarrier approximation to Zak-OTFS, which we refer to as MC-OTFS, has been the focus of most research attention so far \cite{RH1, RH3, br_otfs}. We show that the I/O relation of MC-OTFS is less predictable than that of Zak-OTFS. As the Doppler spread increases, the BER performance of MC-OTFS is inferior to that of Zak-OTFS.
%
Some recent works on OTFS have started focusing on Zak transform based approach \cite{SKM2021, SKM20212, Lampel2022, VSBhat2023}, {\cite{Bondre2022}}. However, none of these 
works investigate the subtle aspect of predictability of the I/O relation in the DD domain. In particular, none emphasizes the important fact that only Zak-OTFS where the information bits are encoded as a discrete quasi-periodic function and filtering is applied through twisted convolution admits a predictable I/O relation if the crystallization condition holds. We feel that this assertion is an important theoretical contribution of this paper.

\textbf{Radar applications}:
We derive the radar ambiguity function for the Zak pulsone and demonstrate that unambiguous delay-Doppler estimation is possible in the crystalline regime, when the delay period is greater than the delay spread of the radar scene, and the Doppler period is greater than the Doppler spread.  We highlight the similarity between the structure of the carrier waveform proposed by Woodward in his influential text \cite{PMWoodward} (a train of narrow TD Gaussian pulses modulated with a broad Gaussian envelope), and the Zak pulsone carrier waveform (a train of narrow impulses modulated by a sinusoid). Note that in radar applications, a signal is modulated onto a carrier waveform (in our case, the Zak pulsone), and that the ambiguity function of the concatenated system depends on both signal and carrier (see \cite{LM2004} and \cite{S2008}).

The paper is organized as follows. Section \ref{secOTFSPredictability} explains how non-predictability in communications results from aliasing in the DD domain, which occurs when one of the channel spreads is greater than the corresponding pulsone period. This geometric perspective leads to a canonical decomposition of the effective channel response filter into a predictable component and a nonpredictable component (the crystalline decomposition). Section \ref{seciorelation} addresses finite bandwidth and duration constraints, expressing the I/O relation for TDM, FDM, and Zak-OTFS in matrix-vector form. Section \ref{secperfcsi} uses BER simulations when the channel is perfectly known, to illustrate that Zak-OTFS is non-fading in the crystalline regime. Section \ref{section5paper2} uses BER simulations when the channel is not known, to illustrate that Zak-OTFS is predictable, and that model-free operation is possible in the crystalline regime. Section \ref{secotfsvariants} uses BER simulations to illustrate that Zak-OTFS is more predictable than MC-OTFS (a widely studied multicarrier approximation) making the case that model-free operation is more possible. Section \ref{radar} derives the radar ambiguity function of the Zak-OTFS carrier waveform, illustrating that unambiguous delay-Doppler estimation is possible in the crystalline regime. Section \ref{conc} presents conclusions.
{
Table-\ref{tabglossary} collects terminology and notation used in the sequel, and provides pointers to additional information.}
{
\begin{table}
\caption{Glossary}
\centering
{
\begin{tabular}{ | c || c |  c | } 
  \hline
   Notation & Description &  Reference\\
   \hline
  $*_{\sigma}$ & Twisted & Footnote $3$ here,   \\ 
   & convolution &  Eq. $(8)$ in \cite{RH1}\\
  \hline
  $\star$ & Linear convolution & \\
  \hline
  ${\mathcal Z}_t$  & Zak transform & Eq. $(4)$ in \cite{PartIpaper} \\
  \hline
  ${\mathcal Z}_t^{-1}$  & Inverse Zak transform & Eq. $(7)$ in \cite{PartIpaper} \\
  \hline
  $\tau_p$, $\nu_p$ & Delay and Doppler period &  Fig.$3$ in \cite{PartIpaper} \\
  & of Zak-OTFS modulation & \\
  \hline
  ${\mathcal D}_0$ & Fundamental DD period & Eq.~$(6)$ in \cite{PartIpaper} \\
  \hline
  Quasi-periodic & Attribute of DD domain &  Eq.~(\ref{eqnqp24}) and Eq.(\ref{paper2_eqn8a}) \\
  & realization of TD signal &  for continuous and \\
  & & discrete DD signals \\
  & & respectively \\
   \hline
  Quasi-periodic & DD domain & Eq.~(15) in \cite{PartIpaper}\\
  DD pulse  & Dirac impulses at & \\
  at $(\tau_a, \nu_a)$ & $(\tau_a + n\tau_p, \nu_a + m \nu_p)$, & \\
  (continuous) & $n,m \in {\mathbb Z}$ & \\
  \hline
  $w_{tx}(\tau,\nu)$ & Transmit pulse shaping filter & Eq.~(\ref{paper2_eqnwtx}), Fig.~\ref{figzakotfspaper2} here\\
   \hline
  $w_{rx}(\tau,\nu)$ & Receive pulse shaping filter & Eq.~(\ref{paper2_eqn229}), Fig.~\ref{figzakotfspaper2} here\\
  \hline
  Zak-OTFS & TD realization of  & ${\mathcal Z}_t^{-1}{\Big (} w_{tx}(\tau, \nu) \,\,  *_{\sigma}  $ \\
  pulsone & filtered quasi-periodic 
  &  $\,\,\,\, \,\,\, \,\,\,\, \, x_{_{\mbox{\scriptsize{dd}}}}(\tau, \nu) {\Big )} $  \\
  & DD pulse $x_{_{\mbox{\scriptsize{dd}}}}(\tau, \nu)$ & \\
  \hline
  $B$ and $T$ & Bandwidth and time duration & \\
  & of modulated TD signal & \\
  \hline
  $M = B \tau_p$ &  Delay domain period & Section \ref{seczakotfsio} here\\
  & of discrete DD signals & \\
  \hline
  $N = T \nu_p$ &  Doppler domain period  & Section \ref{seczakotfsio} here\\
  & of discrete DD signals & \\
   \hline
  Quasi-periodic & $x^{(b)}[k ,l]$ & Eq.~(\ref{paper2_eqn8a1}) here \\
  DD pulse & DD pulse at & \\
  (discrete) & $(k^{(b)}, l^{(b)})$ & \\
  \hline
  $h_{_{\mbox{\scriptsize{phy}}}}(\tau, \nu)$  & DD spreading function &  Eq.$(2)$ in \cite{PartIpaper},\\
  & of underlying channel & Eq.~(\ref{paper2_eqn1}) here \\
  \hline
  $h_{_{\mbox{\scriptsize{eff}}}}(\tau, \nu)$ & Effective continuous DD & Eq.~(\ref{paper2_eqn229}) here\\
  & domain channel filter & \\
  \hline
  $\tau_{_{\mbox{\scriptsize{eff}}}}$ and $\nu_{_{\mbox{\scriptsize{eff}}}}$ & Effective delay and & Eq.~(\ref{paper2_crystallization}) here\\
   & Doppler spread of $h_{_{\mbox{\scriptsize{eff}}}}(\tau, \nu)$ & \\
  \hline
  $h_{_{\mbox{\scriptsize{eff}}}}[k, l]$ & Effective discrete DD & Eq.~(\ref{eqn53_b12}) here\\
  & domain channel filter & \\
  \hline
  Crystallization &  Delay and Doppler periods &  Eq.~(\ref{paper2_crystallization}) here\\
  condition &  exceed respective effective  & \\
  &  channel spreads & \\
  \hline
  Crystalline &  $(\tau_p, \nu_p)$ values which satisfy &  Section \ref{secIntro} here\\
  regime &  crystallization condition &  \\
  \hline
  ${\Lambda}_{_{\mbox{\scriptsize{dd}}}}$ & Zak-OTFS information grid & Eq.~$(27)$ in \cite{PartIpaper} \\
  \hline
  Predictability & Channel response to any & Section \ref{secCrystalline} here\\
  of I/O relation &  input can be predicted & \\
  & from the response to a pulse & \\
  & in modulation domain & \\
  \hline
\end{tabular}}
\label{tabglossary}
\end{table}}

\begin{figure*}
\vspace{-8mm}
\centering
\includegraphics[width=17cm, height=5.5cm]{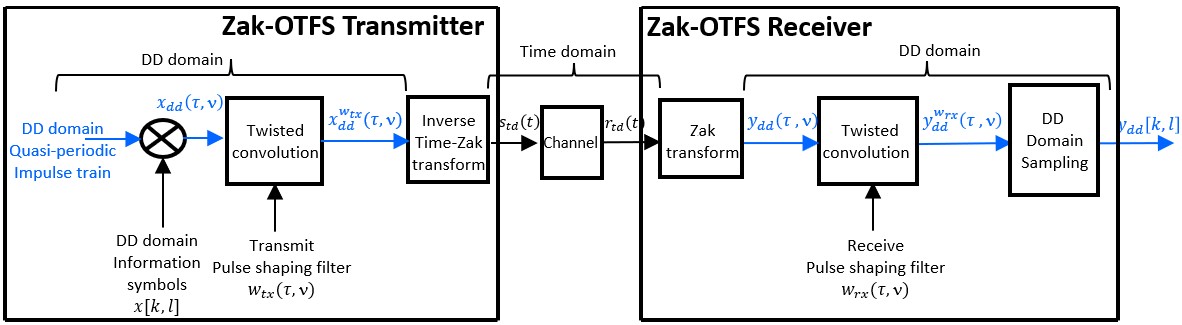}
\caption{Signal processing in Zak-OTFS.}
\label{figzakotfspaper2}
\end{figure*}

\section{The Zak-OTFS Input-Output (I/O) Relation and Predictability in the Crystalline Regime}
\label{secOTFSPredictability}

\subsection{The Zak-OTFS I/O Relation for a Sampled System}
\label{seczakotfsio}
The Zak-OTFS I/O relation (in the absence of AWGN) is presented in Section VI-C of Part I \cite{PartIpaper} (see also Table III in Part I). 
Fig.~\ref{figzakotfspaper2} illustrates Zak-OTFS transceiver processing as explained in the following.

In Zak-OTFS, the frame duration $T$ is roughly an integer multiple
$N$ of the delay period $\tau_p$, and the bandwidth $B$ is roughly an integer multiple $M$ of the Doppler period $\nu_p= 1/\tau_p$.
The information symbols are arranged as a 2-D array $x[k,l]$, $k=0,1,\cdots,M-1$, $l=0,1,\cdots, N-1$ and are encoded as a discrete DD domain information signal which is given by 
\begin{eqnarray}
\label{paper2_eqn8}
x_{_{\mbox{\footnotesize{dd}}}}[k + n M ,l+ m N] & \hspace{-2mm} \Define
x[k,l] \, e^{j 2 \pi n \frac{l}{N}}, \,\,\, m,n \in {\mathbb Z}.
\end{eqnarray}
Being discrete DD domain signal means that $x_{_{\mbox{\footnotesize{dd}}}}[k,l]$ is a quasi-periodic function on the information grid with period $M$ along the delay axis and period $N$ along the Doppler axis, i.e., 
\begin{eqnarray}
\label{paper2_eqn8a}
x_{_{\mbox{\footnotesize{dd}}}}[k+nM, l + mN] & \hspace{-3mm}  = & \hspace{-3mm} x_{_{\mbox{\footnotesize{dd}}}}[k,l] \, e^{j 2 \pi n \frac{l}{N}},
\end{eqnarray}
for any $n,m \in {\mathbb Z}$. The discrete DD domain signal is then lifted to a continuous DD domain signal by means of\footnote{\footnotesize{{We use the same notation $\delta$ for both discrete and continuous Dirac-delta functions, $\delta[\cdot]$ for the discrete case and $\delta(\cdot)$ for the continuous case.}}}
\begin{eqnarray}
\label{paper2_eqnprf37}
x_{_{\mbox{\footnotesize{dd}}}}(\tau, \nu)  & \hspace{-3mm} =  & \hspace{-3mm}  \sum\limits_{k,l \in {\mathbb Z}}  x_{_{\mbox{\footnotesize{dd}}}}[k,l] \, \delta\left(\tau  - k \frac{\tau_p}{M} \right) \delta\left(\nu - l \frac{\nu_p}{N} \right).
\end{eqnarray}
{The encoding in (\ref{paper2_eqn8}) ensures that the continuous DD domain signal $x_{_{\mbox{\footnotesize{dd}}}}(\tau, \nu)$ is \emph{quasi-periodic}, i.e.
for all $n,m \in {\mathbb Z}$}
{\begin{eqnarray}
\label{eqnqp24}
x_{_{\mbox{\footnotesize{dd}}}}(\tau + n \tau_p, \nu + m \nu_p) & = & e^{j 2 \pi n \nu \tau_p} \, x_{_{\mbox{\footnotesize{dd}}}}(\tau, \nu).
\end{eqnarray}}

{ Fig.~\ref{figzakotfspaper2} shows that
twisted convolution of the transmit pulse $w_{tx}(\tau, \nu)$ with $x_{_{\mbox{\footnotesize{dd}}}}(\tau, \nu)$ gives 
the DD domain transmit signal
\begin{eqnarray}
\label{paper2_eqnwtx}
    x_{_{\mbox{\footnotesize{dd}}}}^{w_{tx}}(\tau, \nu) & = &  w_{tx}(\tau, \nu) *_{\sigma} x_{_{\mbox{\footnotesize{dd}}}}(\tau, \nu),
\end{eqnarray}where $*_{\sigma}$ denotes the twisted convolution operation.\footnote{\footnotesize{{$a(\tau, \nu) *_{\sigma} b(\tau, \nu) \Define \iint a(\tau',\nu') \, b(\tau - \tau', \nu - \nu') \, e^{j 2 \pi \nu'(\tau - \tau')} d\tau' \, d\nu'$.}}}
The inverse Zak transform of $x_{_{\mbox{\footnotesize{dd}}}}^{w_{tx}}(\tau, \nu)$ gives the TD realization 
$s_{_{\mbox{\footnotesize{td}}}}(t)$ which is then transmitted. Twisted convolution with the transmit pulse is required for limiting the transmit signal in time and bandwidth.}

The AWGN free part of the received TD signal $r_{_{\mbox{\footnotesize{td}}}}(t)$ is
given by \cite{Bello63}

{
{\vspace{-4mm}
\small
\begin{eqnarray}
\iint h_{_{\mbox{\scriptsize{phy}}}}(\tau,\nu) \, s_{\mbox{\footnotesize{td}}}(t-\tau) \, e^{j 2 \pi \nu (t - \tau)} \, d\tau \, d\nu
\end{eqnarray}\normalsize}where $h_{_{\mbox{\scriptsize{phy}}}}(\tau,\nu)$ is the delay-Doppler representation/spreading function of the underlying physical channel.}

The received TD signal is converted to its DD domain representation $y_{_{\mbox{\footnotesize{dd}}}}(\tau,\nu)$ via the Zak transform.
The channel acts on $x_{_{\mbox{\footnotesize{dd}}}}^{w_{tx}}(\tau, \nu)$ by twisted convolution, so that
$y_{_{\mbox{\footnotesize{dd}}}}(\tau,\nu) = h_{_{\mbox{\scriptsize{phy}}}}(\tau, \nu) *_{\sigma} x_{_{\mbox{\footnotesize{dd}}}}^{w_{tx}}(\tau, \nu)$ (see \cite{PartIpaper}).
After twisted convolution of $y_{_{\mbox{\footnotesize{dd}}}}(\tau,\nu)$ with a receive DD pulse $w_{rx}(\tau, \nu)$ we obtain
\begin{eqnarray}
\label{paper2_eqn229}
y_{_{\mbox{\footnotesize{dd}}}}^{w_{rx}}(\tau, \nu) & = &
w_{rx}(\tau,\nu) \, *_{\sigma} \, y_{_{\mbox{\footnotesize{dd}}}}(\tau, \nu) \nonumber \\
\hspace{-33mm} & \hspace{-30mm} = &  \hspace{-18mm}
w_{rx}(\tau,\nu) \, *_{\sigma} \,  {\Big (} h_{_{\mbox{\scriptsize{phy}}}}(\tau, \nu) \, *_{\sigma} \, x_{_{\mbox{\footnotesize{dd}}}}^{w_{tx}}(\tau, \nu) {\Big )}  \nonumber \\
\hspace{-33mm} & \hspace{-30mm} = &  \hspace{-18mm} w_{rx}(\tau,\nu) \, *_{\sigma} \,  {\Big (} h_{_{\mbox{\scriptsize{phy}}}}(\tau, \nu) \, *_{\sigma} \, {\Big [} 
  w_{tx}(\tau, \nu) *_{\sigma} x_{_{\mbox{\footnotesize{dd}}}}(\tau, \nu) {\Big ]} {\Big )}  \nonumber \\
\hspace{-33mm} & \hspace{-30mm} = &  \hspace{-18mm} \underbrace{{\Big (}  w_{rx}(\tau,\nu) \, *_{\sigma} \, h_{_{\mbox{\scriptsize{phy}}}}(\tau, \nu) \, *_{\sigma} \, w_{tx}(\tau, \nu) {\Big )}}_ {\Define \,\, h_{_{\mbox{\footnotesize{eff}}}}(\tau, \nu)} \, *_{\sigma} \, x_{_{\mbox{\scriptsize{dd}}}}(\tau, \nu),
\end{eqnarray}
where the last step follows from the associativity  of the twisted convolution operation. {To verify associativity of twisted convolution, consider the cascade of two doubly-spread channels with DD spreading functions $h_i (\tau, \nu), i=1,2$,  where the output of channel $1$ is input to channel $2$. This cascade is equivalent to a single doubly-spread channel having DD spreading function $h(\tau, \nu) = h_2(\tau, \nu) *_{\sigma} h_1(\tau, \nu)$ \cite{RH1}. The output of channel $2$ is $y_2(\tau, \nu) = h_2(\tau, \nu) *_{\sigma} {\Big (} h_1(\tau, \nu) *_{\sigma} x_1(\tau, \nu){\Big )}$ where $x_1(\tau, \nu)$ is the input to channel $1$ and ${\Big (} h_1(\tau, \nu) *_{\sigma} x_1(\tau, \nu){\Big )}$ is the output of channel $1$. This expression for $y_2$ must be equal to the expression obtained by considering the channel cascade as a single channel, i.e., $y_2(\tau, \nu) = {\Big (} h_2(\tau, \nu) *_{\sigma} h_1(\tau, \nu) {\Big )} *_{\sigma} x_1(\tau, \nu)$. This proves that $*_{\sigma}$ is associative.}

Equation (\ref{paper2_eqn229}) gives the Zak-OTFS I/O relation in the continuous DD domain. Simply put, the I/O relation states that the output $y_{_{\mbox{\footnotesize{dd}}}}^{w_{rx}}(\tau, \nu)$ is the twisted convolution of the input $ x_{_{\mbox{\footnotesize{dd}}}}(\tau, \nu)$ with the effective continuous DD channel filter $h_{_{\mbox{\scriptsize{eff}}}}(\tau, \nu)$.
Finally, we sample this continuous output signal along the information grid $\Lambda_{_{\mbox{\tiny{dd}}}}$ (see \cite{PartIpaper})
consisting of integer multiples of
$\tau_p/M$ along the delay axis and integer multiples of $\nu_p/N$ along the Doppler axis, to obtain a discrete DD domain output signal
\begin{eqnarray}
\label{paper2_eqn5paper12}
y_{_{\mbox{\footnotesize{dd}}}}[k,l] & \hspace{-3mm} = &  \hspace{-3mm} y_{_{\mbox{\footnotesize{dd}}}}^{w_{rx}}\left(\tau = k \frac{\tau_p}{M}, \nu = l \frac{\nu_p}{N} \right) \nonumber \\
& \hspace{-10mm} = & \hspace{-6mm} \sum\limits_{k', l' \in {\mathbb Z}}  h_{_{\mbox{\scriptsize{eff}}}}[k - k', l - l'] \, x_{_{\mbox{\footnotesize{dd}}}}[k' , l']
\, e^{j 2 \pi \frac{(l - l')}{N} \frac{k'}{M} }  ,
\end{eqnarray}
where $h_{_{\mbox{\scriptsize{eff}}}}[k,l]$ is the discrete effective channel filter, given by sampling the continuous effective channel filter, i.e.,
\begin{eqnarray}
\label{eqn53_b12}
h_{_{\mbox{\scriptsize{eff}}}}[k, l]  & \Define &  h_{_{\mbox{\scriptsize{eff}}}}(\tau,\nu){\Big \vert}_{\left( \tau = \frac{k \tau_p}{M} \,,\, \nu = \frac{l \nu_p}{N} \right)}.
\end{eqnarray}
Equation (\ref{paper2_eqn5paper12}) constitutes the canonical form of the Zak-OTFS I/O relation. It expresses the output signal as a discrete twisted convolution\footnote{\footnotesize{$u[k,l] \, *_{\sigma}  \, v[k,l] = \sum\limits_{k', l' \in {\mathbb Z}} u[k', l'] v[k-k',l-l'] e^{j 2 \pi \frac{k-k'}{M} \frac{l'}{N}}$}} of the discrete effective channel filter and the input signal, i.e.,
\begin{eqnarray}
\label{paper2_eqn764}
\hspace{-5mm} y_{_{\mbox{\footnotesize{dd}}}}[k,l] & \hspace{-5mm} = & \hspace{-6mm} \sum\limits_{k', l' \in {\mathbb Z}}  h_{_{\mbox{\scriptsize{eff}}}}[k', l'] \, x_{_{\mbox{\footnotesize{dd}}}}[k - k' ,l - l']
\, e^{j 2 \pi \frac{(k - k')}{M} \frac{l'}{N} } \nonumber \\  & = & h_{_{\mbox{\scriptsize{eff}}}}[k, l] \, *_{\sigma} \,  x_{_{\mbox{\footnotesize{dd}}}}[k ,l].
\end{eqnarray}
We remark that the R.H.S of (\ref{paper2_eqn5paper12}) is a weighted double sum of terms, where the $(k',l')$-th term is given by 
\begin{eqnarray}
\label{paper2_eqn301}
h_{_{\mbox{\scriptsize{eff}}}}[k,l]   \, *_{\sigma} \,  \left[ \delta[k- k'] \, \delta[l - l'] \right], \,\, k,l \in {\mathbb Z},
\end{eqnarray}
multiplied by the weight $x_{_{\mbox{\footnotesize{dd}}}}[k' ,l']$.

{The Zak-OTFS I/O relation (\ref{paper2_eqn5paper12}) for doubly-spread channels is a generalization of the TDM I/O relation for linear time-invariant (LTI) channels. In an LTI (i.e., delay-only) channel, the match-filtered received continuous TD signal is given by $y(t) = w_{rx}(t) \star h(t) \star w_{tx}(t) \star {\Big (} \sum\limits_{k} x[k] \, \delta(t - k/B) {\Big )} $ where, $h(t)$ is the channel impulse response, $w_{rx}(t)$ and $w_{tx}(t)$ are the transmit and receive filters, $x[k]$ is the discrete information signal and $\star$ denotes linear convolution. Sampling $y(t)$ on the TDM information grid (time instances which are integer multiples of $1/B$) gives the discrete output $y[k] = h[k] \star x[k]$, where $h[k]$ is the discrete effective channel filter, obtained by sampling $w_{rx}(t) \star h(t) \star w_{tx}(t)$ on the TDM information grid.
Similarly, for Zak-OTFS modulation on a doubly-spread channel, $h_{_{\mbox{\scriptsize{eff}}}}[k,l]$, in (\ref{paper2_eqn5paper12}), is the discrete DD domain effective channel filter, $y_{_{\mbox{\footnotesize{dd}}}}[k,l]$ and $x_{_{\mbox{\footnotesize{dd}}}}[k,l]$ are the discrete DD domain channel output and input respectively. This structural similarity between TDM and Zak-OTFS is captured in Fig.~$12$ of \cite{PartIpaper}, and in Tables I and III of \cite{PartIpaper}.}

{An important difference between an LTI and a LTV channel is that the former has eigen-modes/eigen-functions (a pulse in frequency domain, i.e., a TD sinusoid)
whereas the latter does not have eigen-functions. LTV channels however have geometric-modes/geometric-functions. These geometric-modes are Zak-OTFS pulsones which are quasi-periodic pulses in the DD domain and the action of a channel induced delay and Doppler shift is to \emph{geometrically} translate the pulse by the same shifts along the respective domains.}

\subsection{Predictability of the Zak-OTFS I/O relation in the Crystalline Regime}
\label{secCrystalline}
In this section, we show that the Zak-OTFS I/O relation is predictable in the crystalline regime, {i.e., when $(\tau_p, \nu_p)$ ($\tau_p \, \nu_p = 1$) satisfy the crystallization condition}
{\begin{eqnarray}
\label{paper2_crystallization}
\tau_p & > & \tau_{\mbox{\scriptsize{eff}}} \,\,,\,\, \nu_p \, >  \, \nu_{\mbox{\scriptsize{eff}}}
\end{eqnarray}}
{where $\tau_{\mbox{\scriptsize{eff}}}$ and $\nu_{\mbox{\scriptsize{eff}}}$ denote the spread of the support of
$h_{_{\mbox{\scriptsize{eff}}}}(\tau, \nu)$
along the delay and Doppler axis respectively.}

\begin{figure}
\hspace{-2mm}
\includegraphics[width=9cm, height=6.0cm]{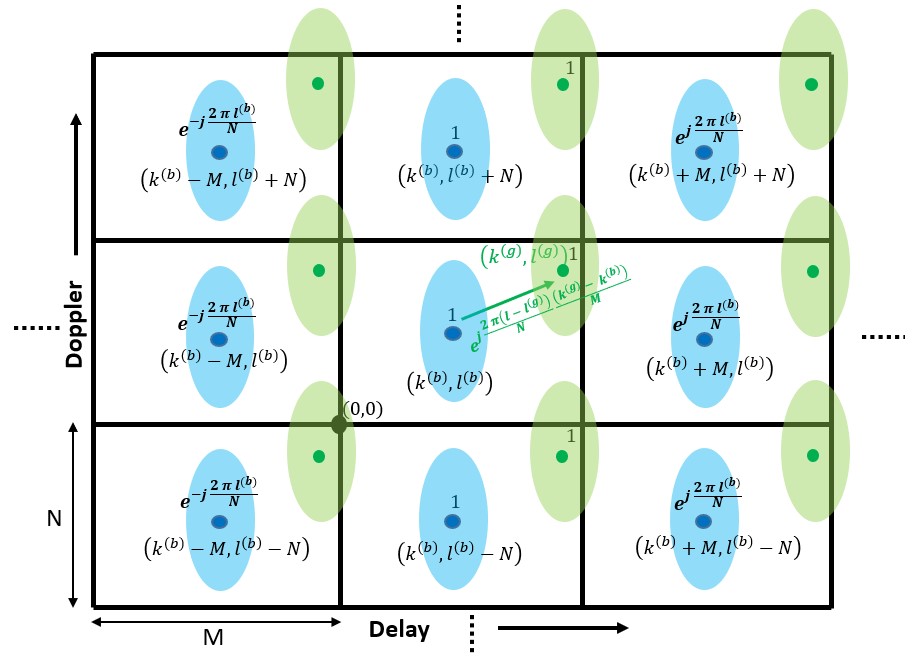}
\caption{Prediction in the crystalline regime. In the crystalline regime, the channel response
to the green DD pilot signal can be predicted from the channel response to the blue DD pilot signal.}
\label{fig1paper2}
\end{figure}

 Specifically, we show that when the crystallization condition holds,
the channel response to a \emph{green} pilot located at $(k^{(g)}, l^{(g)})$ can be accurately estimated from the channel response to a \emph{blue} pilot located at $(k^{(b)}, l^{(b)})$ {(see Fig.~\ref{fig1paper2})}.
{A special case is a delay-only channel (i.e., LTI channel) where we know that the response to a particular impulse input can be used to predict/estimate the channel response to any other impulse input.}

By definition, the blue pilot is a discrete DD domain impulse signal, given by

{\vspace{-4mm}
\small
\begin{eqnarray}
\label{paper2_eqn8a1}
\hspace{-2mm} x^{(b)}[k ,l] & \hspace{-3mm} = & \hspace{-5mm} \sum\limits_{m,n \in {\mathbb Z}} \hspace{-2mm} e^{j 2 \pi \frac{n l^{(b)}}{N}} \delta[k - (k^{(b)} + nM)] \delta[l - (l^{(b)} + mN)].
\end{eqnarray}\normalsize}
Similarly, the green pilot is a discrete DD domain impulse signal, given by

{\vspace{-4mm}
\small
\begin{eqnarray}
\label{paper2_eqn8a1green}
\hspace{-3mm} x^{(g)}[k ,l] & \hspace{-3mm} = & \hspace{-5mm} \sum\limits_{m,n \in {\mathbb Z}} \hspace{-2mm} e^{j 2 \pi \frac{n l^{(g)}}{N}} \delta[k - (k^{(g)} + nM)] \delta[l - (l^{(g)} + mN)].
\end{eqnarray}\normalsize}
The channel response to the blue pilot is given by

{\small
\vspace{-4mm}
\begin{eqnarray}
\label{paper2_eqn8b}
y^{(b)}[k,l] & \hspace{-3mm} = & \hspace{-3mm}  
 h_{_{\mbox{\scriptsize{eff}}}}[k,l] \, *_{\sigma} 
 \, x^{(b)}[k ,l] \nonumber \\
& \hspace{-3mm} = & \hspace{-3mm} \sum\limits_{n,m \in {\mathbb Z}} {\Big [} h_{_{\mbox{\scriptsize{eff}}}}[k - (k^{(b)} + nM), l - (l^{(b)} + mN)] \, \nonumber \\
& & \,\,\, \hspace{6mm} e^{j 2 \pi \frac{n l^{(b)}}{N}} \, e^{j 2 \pi \frac{(l - l^{(b)} - mN)}{N} \frac{(k^{(b)} + n M)}{M} } {\Big ]}.
\end{eqnarray}
\normalsize}

We see from (\ref{paper2_eqn8b}) that the total response to the blue pilot is a sum of local responses to its constituent impulses where the response to the $(n,m)$-th impulse is given by
\begin{eqnarray}
\label{paper2_eqn8d}
y^{(b)}_{_{\mbox{\footnotesize{n,m}}}}[k,l] & \hspace{-3mm} = & \hspace{-3mm} h_{_{\mbox{\scriptsize{eff}}}}[k - (k^{(b)} + nM), l - (l^{(b)} + mN)] \, \nonumber \\
& & \,\,\, \hspace{3mm} e^{j 2 \pi \frac{n l^{(b)}}{N}} \, e^{j 2 \pi \frac{(l - l^{(b)} - mN)}{N} \frac{(k^{(b)} + n M)}{M} }.
\end{eqnarray}
Observe that the $(n,m)$-th response coincides up to multiplicative phases with the effective channel filter shifted by $(k^{(b)} + nM, l^{(b)} + m N)$.

Similarly, the response to the green pilot is a sum of local responses, where the $(n,m)$-th response is given by
\begin{eqnarray}
\label{paper2_eqn8f}
y^{(g)}_{_{\mbox{\footnotesize{n,m}}}}[k,l] & = & {\Big [} h_{_{\mbox{\scriptsize{eff}}}}[k - (k^{(g)} + nM), l - (l^{(g)} + mN)] \, \nonumber \\
& & \,\,\,  e^{j 2 \pi \frac{n l^{(g)}}{N}} \, e^{j 2 \pi \frac{(l - l^{(g)} - mN)}{N} \frac{(k^{(g)} + n M)}{M} } {\Big ]}.
\end{eqnarray}
In Fig.~\ref{fig1paper2}, the response components to the blue (green) pilot are depicted as blue (green) ellipses {(ellipses are used only as visual metaphor and the discussion here is valid for any arbitrary shape of the support set of $h_{_{\mbox{\scriptsize{eff}}}}[k,l]$).} Observe that the $(n,m)$-th green response can always be calculated from the $(n,m)$-th blue response according to the following rule 
\begin{eqnarray}
\label{paper2_eqn8g}
y^{(g)}_{_{\mbox{\footnotesize{n,m}}}}[k,l] & = & y^{(b)}_{_{\mbox{\footnotesize{n,m}}}}[k - (k^{(g)} - k^{(b)}),l -(l^{(g)} - l^{(b)})] \nonumber \\
& & \hspace{-4mm} e^{j 2 \pi n \frac{(l^{(g)} - l^{(b)})}{N}} \, e^{j2 \pi \frac{(l - l^{(g)} - mN)}{N} \frac{(k^{(g)} - k^{(b)})}{M}}.
\end{eqnarray}
where we use (\ref{paper2_eqn8d}) and (\ref{paper2_eqn8f}).\footnote{\footnotesize{To obtain the $(n,m)$-th response component to the green pilot from that of the blue pilot, we first align supports by translating the blue ellipses by $(k^{(g)} - k^{(b)})$ in delay, and by $(l^{(g)} - l^{(b)})$ in Doppler. We then multiply by a deterministic factor that is independent of the channel.}} The problem is that the individual local responses are super-imposed at the receiver and, in general, cannot be separated from one another. However, when the crystallization conditions hold, the blue (green) ellipses do not overlap with each other which means that the local responses do not interact. As a consequence, in this situation. the total green response can be predicted from the total blue response using the local formula (\ref{paper2_eqn8g}). 

An important consequence of predictability is that the energy profile of the local responses is flat, that is, independent of the position of the pilot, due to the fact that 

{\vspace{-4mm}
\small
\begin{eqnarray}
\label{paper2_nofading1}
\left\vert y^{(g)}_{_{\mbox{\footnotesize{n,m}}}}[k,l] \right\vert = \left\vert y^{(b)}_{_{\mbox{\footnotesize{n,m}}}}[k - (k^{(g)} - k^{(b)}),l -(l^{(g)} - l^{(b)})] \right\vert, 
\end{eqnarray}\normalsize}

In other words, in the crystalline regime, the Zak-OTFS I/O
relation is \emph{non-fading}.

\begin{figure}
\hspace{-7mm}
\includegraphics[width=10cm, height=5.0cm]{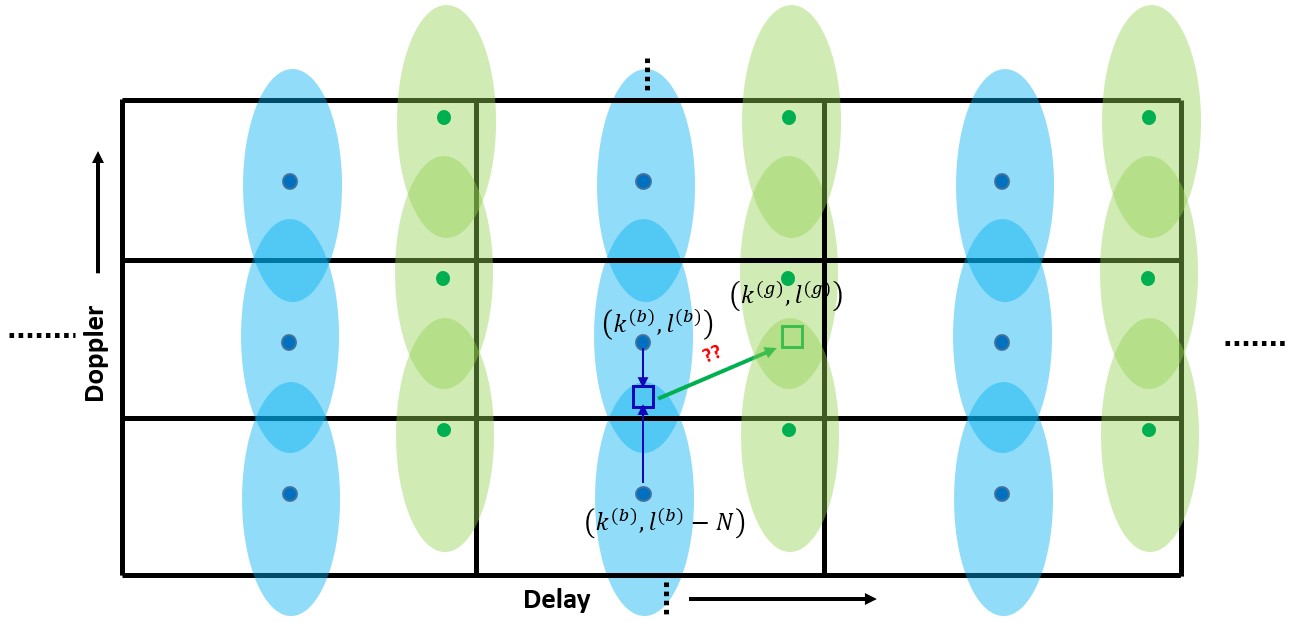}
\caption{Non-crystalline regime: Partially predictable Zak-OTFS I/O relation. The
response to the green DD pilot signal can be predicted from the response to the blue DD pilot signal only for the part of the response which is free from any overlap/aliasing/self-interaction.}
\label{fig2paper2}
\end{figure}

\subsection{Non-predictability of the Zak-OTFS I/O relation in the Non-crystalline Regime}
\label{secNoncrystalline}
When the crystallization conditions fail to hold the local channel responses interact (the ellipses overlap with one another), a phenomenon we refer to as DD domain aliasing. We now explain how non-predictability arises from DD domain aliasing.

We illustrate the aliasing phenomenon through an example. We again look to predict the response to a green pilot from the response to a blue pilot. However, we now assume that the Doppler spread of the effective channel filter is greater than the Doppler period. Under this assumption, local responses interact. Specifically, in our example the $(n,m)$-th response interacts with the $(n,m-1)$-th response. The interaction is depicted in Fig.~\ref{fig2paper2} as an overlap between ellipses. 

We consider a generic point $(k,l)$ residing in the overlap of the $(0,0)$ and $(0,-1)$ blue ellipses, depicted in Fig.~\ref{fig2paper2} as a square with a blue border. The response received at this point is the super-position of two local responses $y^{(b)}_{_{\mbox{\footnotesize{0,0}}}}[k,l] + y^{(b)}_{_{\mbox{\footnotesize{0,-1}}}}[k,l]$, which is equal to

{\vspace{-4mm}
\small
\begin{eqnarray}
\label{paper2_eqn8l}
h_{_{\mbox{\scriptsize{eff}}}}[\Delta k, \Delta l] e^{j 2 \pi \frac{\Delta l}{N} \frac{k^{(b)}}{M}} + h_{_{\mbox{\scriptsize{eff}}}}[\Delta k, \Delta l + N] e^{j 2 \pi \frac{\Delta l + N}{N} \frac{k^{(b)}}{M}},
\end{eqnarray}
\normalsize}where $\Delta k = k - k^{(b)}$ and $\Delta l = l - l^{(b)}$. We consider the parallel point $(k',l')$ residing in the overlap of the $(0,0)$ and $(0,-1)$ green ellipses, depicted in Fig.~\ref{fig2paper2} as a square with a green border. Here, $k'=k+(k^{(g)} - k^{(b)})$ and $l'=l+(l^{(g)} - l^{b)})$. The response at this point is a super-position of two local responses $y^{(g)}_{_{\mbox{\footnotesize{0,0}}}}[k',l'] + y^{(g)}_{_{\mbox{\footnotesize{0,-1}}}}[k',l']$, which is explicitly equal to

{\vspace{-4mm}
\small
\begin{eqnarray}
\label{paper2_eqn8m}
h_{_{\mbox{\scriptsize{eff}}}}[\Delta k, \Delta l] e^{j 2 \pi \frac{\Delta l}{N} \frac{k^{(g)}}{M}} + h_{_{\mbox{\scriptsize{eff}}}}[\Delta k, \Delta l + N] e^{j 2 \pi \frac{\Delta l + N}{N} \frac{k^{(g)}}{M}}.
\end{eqnarray}
\normalsize}

We see that unless both the individual local terms in (\ref{paper2_eqn8l}) are separately known, the green response in (\ref{paper2_eqn8m}) cannot be predicted from the total blue response. 

A consequence of non-predictability is that the energy profile of the received response depends on the DD location of the pilot signal. In our example, the energy profile of the received response to the blue pilot will, in general, be different than that for the green pilot. This is because the linear combination (\ref{paper2_eqn8l}) can be different from (\ref{paper2_eqn8m}), depending on the values of the phase coefficients in each case. 

\subsection{The Crystalline Decomposition} 
\label{secCrysdecomp}
The previous discussion reveals that the channel response is unpredictable at points inside the overlap region between interacting ellipses. Outside this region, the response is in fact predictable. This observation suggests a decomposition of the effective channel filter into a predictable and a non-predictable component. 

To see this, we consider a pilot at the origin. The channel response to this pilot is given by
\begin{eqnarray}
\label{paper2_ddpilotorigin}
h_{_{\mbox{\scriptsize{eff}}}}[k, l]  +  \hspace{-5mm}\sum\limits_{n,m \in {\mathbb Z}, (n,m) \ne (0,0)} h_{_{\mbox{\scriptsize{eff}}}}[k - nM, l - mN] \, e^{j 2 \pi \frac{n l}{N}}
\end{eqnarray}
where the first term is the local response to the impulse at the origin and the remaining terms are local responses to other impulses. The main observation is that if $(k,l)$ satisfies $h_{_{\mbox{\scriptsize{eff}}}}[k, l] \neq 0$ and $h_{_{\mbox{\scriptsize{eff}}}}[k - nM, l - mN] \neq 0$ for some $(n,m) \ne (0,0)$, then the response at $(k,l)$ is a superposition of at least two local responses -  the local response to the $(0,0)$-th impulse and the local response to the $(n,m)$-th impulse, and, consequently, the effective channel filter tap at $(k,l)$ cannot be estimated from the received response. We refer to such taps as  \emph{non-predictable} taps. Alternatively, it is easy to see that all other taps in the support can be estimated from the channel response. We refer to such taps as \emph{predictable} taps. We denote the support set of all predictable taps by ${\mathcal P}$ and the complement support set of all non-predictable taps by ${\mathcal P}^c$. 

Recall that given a set ${\mathcal S}$, the indicator function ${\chi}_{_{\mathcal S}}[k,l]$ takes the value $1$ for $(k,l) \in {\mathcal S}$, and the value $0$ otherwise. We define the \emph{crystalline decomposition} of the effective channel filter to be

\begin{eqnarray}
h_{_{\mbox{\scriptsize{eff}}}}[k,l] & = & {\chi_{_{{\mathcal P}}}}[k,l] 
 \, h_{_{\mbox{\scriptsize{eff}}}}[k,l] \,\, + \,\,   {\chi_{_{{\mathcal P}^c}}}[k,l] 
 \, h_{_{\mbox{\scriptsize{eff}}}}[k,l],
\end{eqnarray}
where the first term is the predictable component and the second term is the non-predictable component. {A mathematically exact definition of the crystallization condition is that the non-predictable component ${{\mathcal P}}^c$ vanishes, and this condition is weaker than (\ref{paper2_crystallization}). However, we shall focus on the stricter condition given by (\ref{paper2_crystallization}), since it relates directly to delay and Doppler spread, channel characteristics that are important to wireless systems engineering.}

\underline{{Remark:}}
{We can view the inability to predict the Zak-OTFS I/O relation from a single pilot as the infeasibility of inverting a certain linear system. We send a quasi-periodic pilot at the origin, then make $MN$ measurements by sampling the channel DD response on the information grid ($M$ taps along the delay axis and $N$ taps along the Doppler axis). Predictability requires that we are able to calculate the effective channel taps $h_{_{\mbox{\scriptsize{eff}}}}[k, l]$ from the $MN$ measurements. This is not possible, for example, when the number of taps to be estimated exceeds the number of measurements.}

\subsection{{Identification of Linear Time-varying Channels}}
\label{paper2_identifyltv}

{Acquisition of the I/O relation is equivalent to channel identification. Bello \cite{Bello1969} conjectured that a LTV channel is identifiable if and only if the channel is \emph{under-spread}, i.e. the support area $A$ of the channel delay-Doppler (DD) spreading function is less than one.
This conjecture was proved in \cite{Pfander2006} and extended in \cite{Bolcskei2013}. We note that the method developed in \cite{Bolcskei2013} is to use a weighted train of Dirac impulses as a probe signal, then to use the Zak transform at the receiver, to derive a system of linear equations that specifies the channel.}

{The channel estimation algorithm proposed in \cite{Bello1969, Pfander2006, Bolcskei2013} involves solving a linear system of equations with a large number of unknown channel parameters (roughly equal to $B \, T \, A$ where $B \, T$ is the time-bandwidth product of the communication system and $A$ is the support area of the channel DD spreading function). This requires inversion of large matrices which makes practical implementation \emph{challenging}.}

{In contrast,
we use a single pulsone as a probing signal, for example, the TD realization of a quasi-periodic DD pulse at the origin. The channel response to such a probing/pilot signal is given by (\ref{paper2_ddpilotorigin}). When the crystallization conditions in (\ref{paper2_crystallization}) hold, the terms $h_{_{\mbox{\scriptsize{eff}}}}[k - nM, l - mN] \, e^{j 2 \pi \frac{n l}{N}}$ ($(n,m) \ne (0,0)$) in (\ref{paper2_ddpilotorigin}) have support which does not overlap with the support of the effective channel filter $h_{_{\mbox{\scriptsize{eff}}}}[k,l]$ and therefore its taps can be read-off directly from the received response. Therefore, probing a linear time-varying channel with a Zak-OTFS pulsone significantly simplifies acquisition of the I/O relation in comparison to \cite{Bello1969, Pfander2006, Bolcskei2013} where other probing signals are used. The under-spread condition is an intrinsic property of the channel, whereas the crystallization condition is a property of the channel in relation to the $(\tau_p, \nu_p)$ parameters of Zak-OTFS modulation. Note that we use the two-dimensional period lattice (with generators $(\tau_p, 0)$ and $(0, \nu_p)$) to define Zak-OTFS modulation, and that it is the combination of underspread channel and Zak-OTFS modulation that minimizes the complexity of channel estimation.}

{If we ignore modulation and simply focus on channel support, the crystallization condition (\ref{paper2_crystallization}) is \emph{not the same} as the channel under-spread condition $A < 1$. The crystallization condition requires, the delay spread to be less than the delay period ($\tau_p$) and the Doppler spread to be less than the Doppler period ($\nu_p$). Since $\tau_p \times \nu_p = 1$, the crystallization condition is more \emph{restrictive} than the channel under-spread condition, i.e., there are channels which are under-spread ($A < 1$) but for which it is not possible to choose $(\tau_p, \nu_p)$ ($\tau_p \nu_p = 1$) such that the delay and Doppler spreads are less than the respective periods. Being more restrictive, however allows \emph{simple} channel estimation (as discussed above)}.

\subsection{Error in prediction of the Zak-OTFS I/O relation}
In this section we analyze the prediction accuracy for different
choices of the delay-Doppler period and shaping filter. The analysis is carried through simulations of a simple two-path channel, where, the first path has a delay of $\tau_1 = 0 \, \mu s$ and a Doppler shift of $\nu_1 = 815$ Hz and the second path
has a delay of $\tau_2 = 5 \mu s$ and a Doppler shift of $\nu_2 = -815$ Hz. The normalized channel gains for the two paths are $h_1 = h_2 = 1/\sqrt{2}$. The system bandwidth is $B = 0.96$ MHz and the frame duration is $T=1.6$ ms.

We consider two choices for the transmit/receive shaping filters. The first choice is a sinc filter
\begin{eqnarray}
w_{rx}(\tau,\nu) = w_{tx}(\tau,\nu) = \sqrt{B T} \, sinc(B \tau) \, sinc(\nu T).
\end{eqnarray}

Observe that the slow decay of the sinc function amplifies DD domain aliasing, thereby, reducing predictability. 

The second choice is a root raised cosine (RRC) filter
\begin{eqnarray}
\label{rrcpulse_eqn1}
w_{rx}(\tau,\nu) = w_{tx}(\tau,\nu) = \sqrt{BT} \, rrc_{_{\beta_{\tau}}}( B \tau ) \,  rrc_{_{\beta_{\nu}}}( T \nu ),
\end{eqnarray}
where the function $rrc_{\beta}(\cdot)$ for $0 \leq \beta \leq 1$ is given by \cite{SHDigcomm}
\begin{eqnarray}
\label{rrceqn1}
rrc_{_{\beta}}(x) & \hspace{-3mm} = & \hspace{-3mm} \frac{\sin(\pi x (1 - \beta)) + 4 \beta x \cos(\pi x (1 + \beta))}{\pi x \left( 1 -( 4 \beta x )^2 \right)}.
\end{eqnarray}

Observe that the RRC function has faster decay than the sinc function, causing less DD domain aliasing, thereby increasing predictability. However, reduced aliasing comes at the cost of an increase in frame duration and bandwidth. In the simulation we use $\beta_{\tau} = 0.1$ and $\beta_{\nu} = 0.2$, corresponding to a $10 \, \%$ increase in bandwidth, and a
$20 \, \%$ increase in duration.

We normalize the energy of the transmit/receive filters by setting
\begin{eqnarray}
\iint \vert w_{tx}(\tau, \nu) \vert^2 \, d\tau d\nu = \iint \vert w_{rx}(\tau, \nu) \vert^2 \, d\tau d\nu = 1.
\end{eqnarray}

We now use a pilot signal at $(k^{(b)},l^{(b)}) = \left( \frac{M}{2}, \frac{N}{2} \right)$
to estimate the effective channel filter $h_{_{\mbox{\scriptsize{eff}}}}[k,l]$. Recall, that
in the crystalline regime, local responses do not interact, therefore, the total response in the fundamental period coincides with the $(0,0)$-th local response $y^{(b)}_{0,0}[k,l]$ which is given by
\begin{eqnarray}
y^{(b)}[k,l] = h_{_{\mbox{\scriptsize{eff}}}}[k - M/2,l - N/2] \, e^{j \pi \frac{(l - N/2)}{N} },
\end{eqnarray}
for $0 \leq k < M$ and $0 \leq l < N$. From this we conclude
\begin{eqnarray}
\label{paper2_eqn8k}
h_{_{\mbox{\scriptsize{eff}}}}[k,l] =  y^{(b)}[k+ M/2,l + N/2 ]  \,  e^{- j \pi \frac{l}{N} },
\end{eqnarray}
for $-\frac{M}{2} \leq k < \frac{M}{2}$ and $-\frac{N}{2} \leq l < \frac{N}{2}$. We set the channel filter estimate to be

{\vspace{-4mm}
\small
\begin{eqnarray}
\label{paper2_predicteqn1}
{\widehat h}_{_{\mbox{\scriptsize{eff}}}}[k,l] & =
\begin{cases}
{y}^{(b)} \left[ k+\frac{M}{2},l+\frac{N}{2} \right] \, e^{-j \pi \frac{l}{N}}&, -\frac{M}{2} \leq k < \frac{M}{2}\\
& \,\, -\frac{N}{2} \leq l < \frac{N}{2}\\
0 &, \mbox{\small{otherwise.}} \\
\end{cases}.
\end{eqnarray}\normalsize}

Using (\ref{paper2_predicteqn1}), we can predict the effective channel response to any
pilot. The predicted channel response to a pilot located at $(k^{(g)}, l^{(g)})$ is given by

{\small
\vspace{-4mm}
\begin{eqnarray}
\label{paper2_predicteqn24}
\widehat{y}^{(g)}[k,l] & \hspace{-3mm} = & \hspace{-3mm}
{\widehat h}_{_{\mbox{\scriptsize{eff}}}}[k,l] \, *_{\sigma}  x^{(g)}[k,l].
\end{eqnarray}
\normalsize}

The predicted  response should be compared to the true channel response

{\small
\vspace{-4mm}
\begin{eqnarray}
\label{paper2_predicteqn2}
y^{(g)}[k,l] & \hspace{-3mm} = & \hspace{-3mm}
{h}_{_{\mbox{\scriptsize{eff}}}}[k,l] \, *_{\sigma}  x^{(g)}[k,l].
\end{eqnarray}
\normalsize}
\begin{figure}
\vspace{-4mm}
\centering
\includegraphics[width=9cm, height=5.5cm]{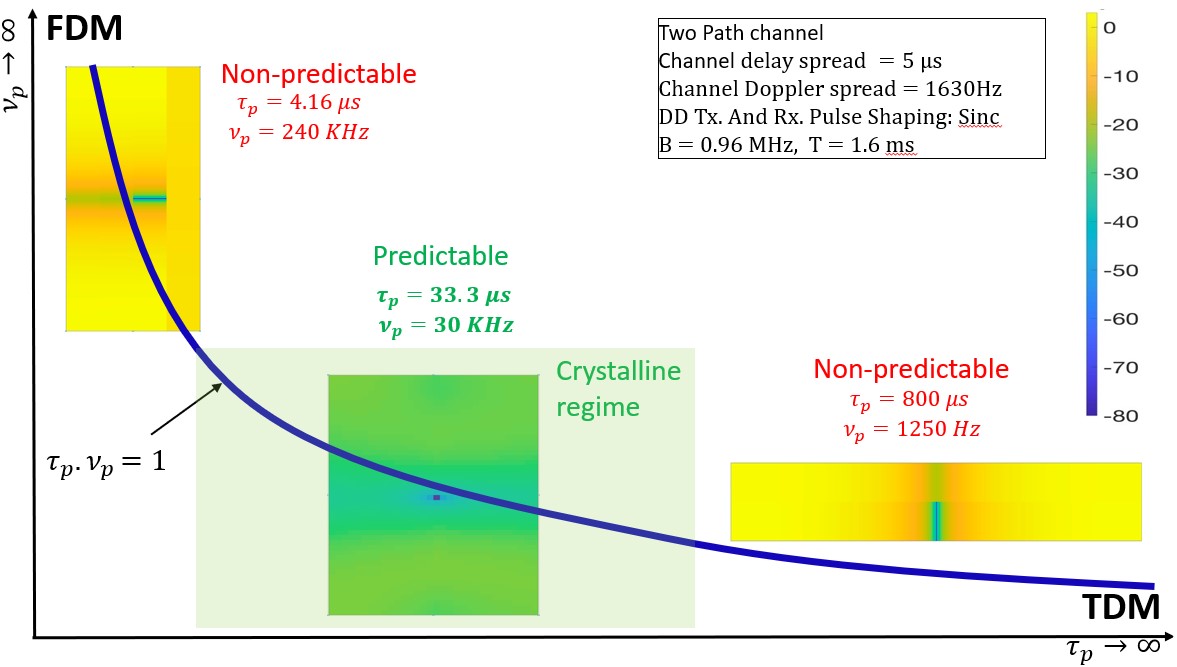}
\caption{Heatmap showing Relative Prediction Error (RPE), in dB, as a function of delay (horizontal axis), and Doppler (vertical axis) with sinc pulse shaping filters. RPE
is significantly smaller in the crystalline regime when compared to that in the non-crystalline regime.}
\label{fig3paper2}
\end{figure}
The relative prediction error is given by
\begin{eqnarray}
\label{paper2_ek1l1eqn}
E(k^{(g)}, l^{(g)}) &  \Define & \frac{\sum\limits_{k=0}^{M-1}\sum\limits_{l=0}^{N-1} {\Big \vert} \widehat{y}^{(g)}[k,l]  -  y^{(g)}[k,l]{\Big \vert}^2}{\sum\limits_{k=0}^{M-1}\sum\limits_{l=0}^{N-1} {\Big \vert} y^{(g)}[k,l]{\Big \vert}^2}.
\end{eqnarray}

\begin{figure}
\vspace{-4mm}
\centering
\includegraphics[width=9cm, height=5.5cm]{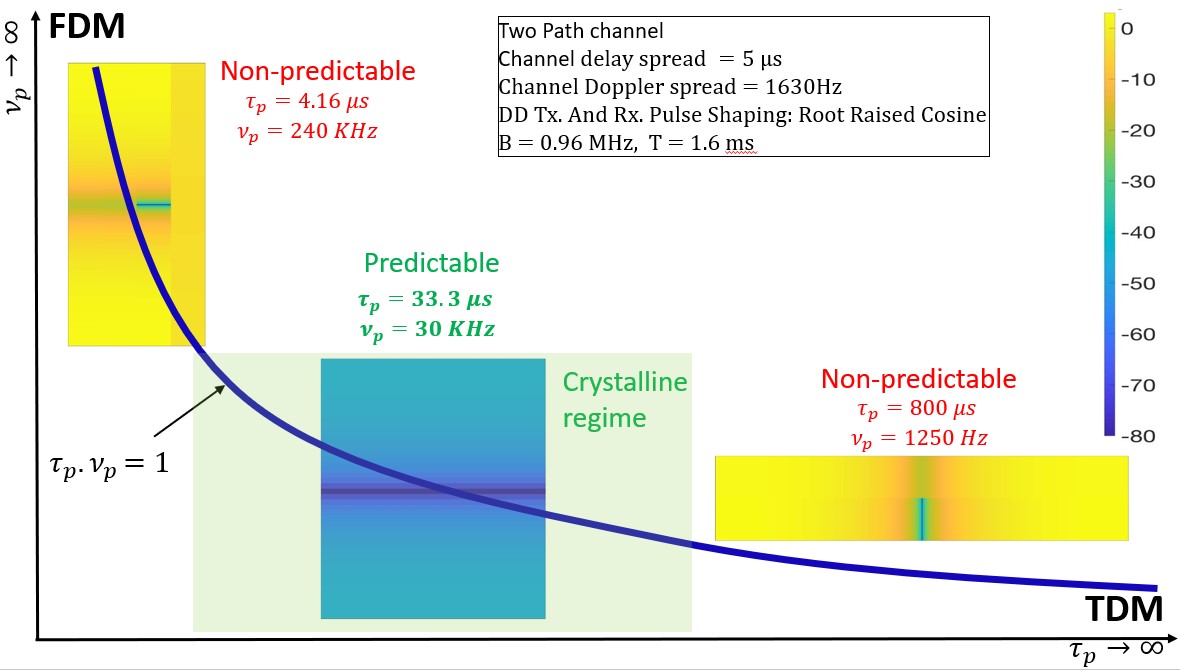}
\caption{Heatmap showing Relative Prediction Error (RPE), in dB, as a function of delay (horizontal axis), and Doppler (vertical axis) with root raised cosine (RRC) pulse shaping filters.
In the crystalline regime, RRC pulse results in significantly smaller RPE when compared to sinc pulses, although at the cost of a higher OTFS frame duration and bandwidth.}
\label{fig3paper2_rc}
\end{figure}

Figs.~\ref{fig3paper2} and \ref{fig3paper2_rc} depict the relative prediction error as a two-dimensional heat-map for three different points on the period curve: the points $\nu_p = 1.25$ KHz, representing a TDM approximation, $\nu_p = 30$ KHz representing a point in the crystalline regime and $\nu_p=240$ KHz, representing an FDM approximation.\footnote{\footnotesize{{In \cite{PartIpaper} it was shown that Zak-OTFS is a family of modulations parameterized by $(\tau_p, \nu_p)$. As the Doppler period $\nu_p$ tends to infinity and the delay period $\tau_p = 1/\nu_p$ tends to zero, the proposed Zak-OTFS modulation behaves like FDM, i.e., Zak-OTFS is approximately FDM. Similarly, when the delay period tends to infinity and the Doppler period tends to zero, the proposed Zak-OTFS is approximately TDM.}}} Fig.~\ref{fig3paper2} assumes sinc transmit/receive shaping filters, whereas Fig.~\ref{fig3paper2_rc} assumes root raised cosine transmit/receive filters.

When $\nu_p = 1.25$ KHz ($\tau_p = 800 \, \mu s$), the Doppler period is smaller than the channel Doppler spread ($1.63$ KHz), causing aliasing along the Doppler dimension which, in turns, creates non-predictability and fading along delay. In this situation the relative prediction error is minimized at the pilot location and it increases rapidly towards $0$ dB as we move away along the delay axis.

When $\nu_p = 240$ KHz, ($\tau_p = 4.16 \, \mu s$), the delay period is smaller than the channel delay spread ($5 \, \mu s$) causing aliasing along along the delay dimension, which, in turns, creates non-predictability and fading along Doppler. In this situation the relative prediction error is minimized at the pilot location, and it now increases rapidly towards $0$ dB as we move away from the pilot location along the Doppler axis.

Finally, when $\nu_p = 30$ KHz, ($\tau_p = 33.3 \, \mu s$), the crystallization condition holds and predictability is maintained. In this situation, the relative prediction error is small. 

When we compare Fig.~\ref{fig3paper2} to Fig.~\ref{fig3paper2_rc}, we see that the choice of transmit/receive shaping filter makes a significant difference. Using root raised cosine filters lead to a relative prediction error of roughly $-50$ dB, whereas using sinc filters lead to a relative prediction error of roughly $-20$ dB. 

\begin{figure}
\includegraphics[width=8cm, height=6.0cm]{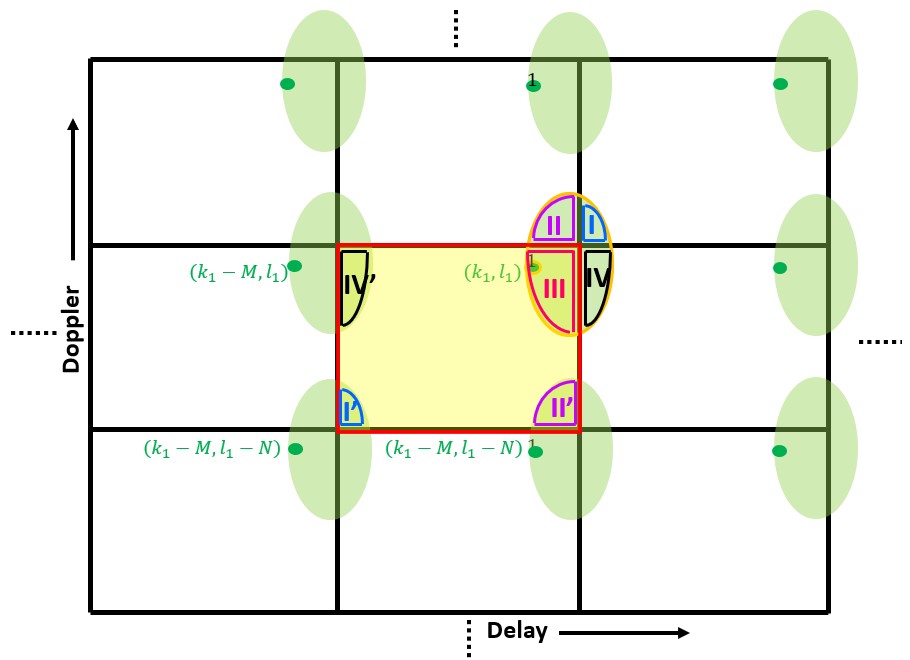}
\caption{Exploiting Quasi-periodicity: Generating complete response from response received in the fundamental DD period.}
\label{fig4paper2}
\end{figure}

\section{Matrix-vector description of the I/O relation}
\label{seciorelation}
{Recall from \cite{Bello63} that for any modulation (TDM/FDM/Zak-OTFS) the channel equation relating the received and the transmitted TD signals is given by }
{
{\vspace{-2mm}
\small
\begin{eqnarray}
\label{paper2_eqn1}
r_{\mbox{\footnotesize{td}}}(t) = \iint h_{_{\mbox{\scriptsize{phy}}}}(\tau,\nu) \, s_{\mbox{\footnotesize{td}}}(t-\tau) \, e^{j 2 \pi \nu (t - \tau)} \, d\tau \, d\nu \, + \, n_{\mbox{\footnotesize{td}}}(t),
\end{eqnarray}\normalsize}where $h_{_{\mbox{\scriptsize{phy}}}}(\tau,\nu)$ is the delay-Doppler representation/spreading function of the underlying physical channel and $n_{\mbox{\footnotesize{td}}}(t)$ is the AWGN noise term. Although there are other equivalent channel representations (e.g., the time-delay representation $h_{_{\mbox{\scriptsize{phy}}}}(t, \tau)$), we consider the DD representation in this paper since the resulting Zak-OTFS I/O relation in the DD domain is described simply in terms of twisted convolution of the DD domain channel input with a discrete DD domain channel filter $h_{_{\mbox{\scriptsize{eff}}}}[k,l]$.
This filter can be acquired efficiently (using a single pulsone probe signal), making it possible, in the crystalline regime, to predict the channel response to any Zak-OTFS input (see (\ref{paper2_eqn764}), Section \ref{secCrystalline} and Section \ref{paper2_identifyltv}).}

In this section we study the I/O relation induced by (\ref{paper2_eqn1}) for Zak-OTFS, TDM and FDM and establish a matrix formulation in each case.

\subsection{Zak-OTFS}
\label{subseczakotfsmatrixIO}
Direct calculation reveals that the DD domain I/O relation induced by (\ref{paper2_eqn1}) is given by

{\vspace{-4mm}
\small
\begin{eqnarray}
\label{paper2_eqn7b}
y_{_{\mbox{\footnotesize{dd}}}}[k,l] 
& \hspace{-3mm} = & \hspace{-3mm} \sum\limits_{k', l' \in {\mathbb Z}}  h_{_{\mbox{\scriptsize{eff}}}}[k - k', l - l'] \, x_{_{\mbox{\footnotesize{dd}}}}[k',l']
\, e^{j 2 \pi \frac{(l - l')}{N} \frac{k'}{M}}  \nonumber \\
& & \hspace{10mm} \, + \, n_{_{\mbox{\footnotesize{dd}}}}[k,l],
\end{eqnarray}\normalsize}where $h_{_{\mbox{\scriptsize{eff}}}}[k, l]$ is the effective DD domain channel filter and $n_{_{\mbox{\footnotesize{dd}}}}[k,l]$ is the discrete noise term obtained by sampling 
\begin{eqnarray*}
n_{_{\mbox{\footnotesize{dd}}}}(\tau, \nu) = w_{rx}(\tau, \nu) \, *_{\sigma} \, {\mathcal Z}_t(n_{_{\mbox{\footnotesize{td}}}}(t)).
\end{eqnarray*}
Here ${\mathcal Z}_t(\cdot)$ denotes the time-Zak transform (see \cite{PartIpaper}).

\underline{{Remark:}}
{The Zak-OTFS I/O relation is valid, whether the crystallization condition is satisfied or not. This is because, the channel action (i.e. twisted convolution in the DD domain) is independent of whether the delay and Doppler period parameters of Zak-OTFS modulation are greater than the respective channel spreads.
This is explained in Part I \cite{PartIpaper}, but it may not be clear from a subsequent paper \cite{Bondre2022} which only considers the I/O relation given by twisted convolution for under-spread channels.
}

Both sequences $x_{_{\mbox{\footnotesize{dd}}}}[k ,l]$ and $y_{_{\mbox{\footnotesize{dd}}}}[k ,l]$ {are quasi-periodic irrespective of whether the crystallization condition is satisfied or not.}\footnote{\footnotesize{{This is because quasi-periodicity is an intrinsic property of the Zak transform (i.e., the Zak transform of an TD signal is quasi-periodic) which is independent of the channel.}}} 

Hence, both sequences can always be reconstructed from samples within the fundamental period ($0 \leq k < M, 0 \leq l < N$), {irrespective of whether we operate in the crystalline regime or not.} Fig.~\ref{fig4paper2} illustrates how to reconstruct the local response to a DD domain pilot supported over the union of regions I, II, III and IV, from the received samples supported over the union of regions I', II', III and IV' within the fundamental period.

{As the received DD domain signal $y_{_{\mbox{\footnotesize{dd}}}}[k ,l]$
is quasi-periodic with period $M$ and $N$ along the delay and Doppler axis respectively}, the $MN$ received samples in the fundamental period constitute a sufficient statistic for the transmitted information symbols $x[k,l]$. Hence, the I/O relation (\ref{paper2_eqn7b}) can be reduced to a finite matrix form relating the $MN \times 1$ vector of received samples $y_{_{\mbox{\footnotesize{dd}}}}[k,l]$ to the $MN \times 1$ vector of transmitted symbols $x[k,l]$, where on both sides $k=0,1,\cdots,M-1$ and  $l=0,1,\cdots, N-1$. In more detail, define the $MN \times 1$ vectors
\begin{eqnarray}
\label{paper2eqn192}
\left({{\bf y}}_{_{\mbox{\footnotesize{dd}}}}\right)_{kM+l+1} & = &  {{ y}}_{_{\mbox{\footnotesize{dd}}}}[k,l] \nonumber \\
\left({{\bf x}}_{_{\mbox{\footnotesize{dd}}}}\right)_{kM+l+1} & = &  {{x}}_{_{\mbox{\footnotesize{dd}}}}[k,l] \nonumber \\
\,\,\, \left({{\bf n}}_{_{\mbox{\footnotesize{dd}}}}\right)_{kM+l+1} & = &  {{n}}_{_{\mbox{\footnotesize{dd}}}}[k,l], 
\end{eqnarray}
In addition, define the $MN \times MN$ matrix 
\begin{eqnarray}
\label{paper2eqn193}
\left( {\bf H}_{_{\mbox{\footnotesize{dd}}}} \right)_{k'N + l'+1, kN+l+1} & \hspace{-3mm}  = & \hspace{-3mm}  {H}_{_{\mbox{\footnotesize{dd}}}}[k'N + l', kN+l]
\end{eqnarray}
where the right hand side is given by (\ref{paper2_eqn12}) (see top of next page).
\begin{figure*}
\vspace{-9mm}
\begin{eqnarray}
\label{paper2_eqn12}
{ H}_{_{\mbox{\footnotesize{dd}}}}[k'N + l' , k N + l ]  & = & \sum\limits_{n=-\infty}^{\infty} \sum\limits_{m = -\infty}^{\infty} h_{_{\mbox{\scriptsize{eff}}}}[k' - k -nM, l' - l -mN] \, e^{j 2 \pi n l/N} e^{j2 \pi \frac{l' - l - mN}{N} \frac{k + nM}{M}}, \nonumber \\
& & k',k=0,1,\cdots, M-1, \,\, l',l=0,1,\cdots, N-1.
\end{eqnarray}
\end{figure*}
The matrix formulation of the DD domain I/O relation (\ref{paper2_eqn7b}) is given by
\begin{eqnarray}
\label{paper2_eqnyddhdd}
{{\bf y}}_{_{\mbox{\footnotesize{dd}}}}  & = & {{\bf H}}_{_{\mbox{\footnotesize{dd}}}}
{{\bf x}}_{_{\mbox{\footnotesize{dd}}}} \, + \, {{\bf n}}_{_{\mbox{\footnotesize{dd}}}}.
\end{eqnarray}
{Since the I/O relation in (\ref{paper2_eqn7b}) is valid irrespective of whether we operate in the crystalline regime or not, the same holds for the corresponding matrix formulation.}

\subsection{TDM}
\label{subsectdmmatrixIO}
Direct calculation reveals that the {discrete} time-domain I/O relation induced from (\ref{paper2_eqn1}), is given by (see Equation ($24$) in Part I) 

{\vspace{-4mm}

\begin{eqnarray}
\label{paper2_eqn2}
y_{\mbox{\footnotesize{td}}}[k] & = & \sum\limits_{k' \in {\mathbb Z}} x_{\mbox{\footnotesize{td}}}[k'] \, h_{_{\mbox{\footnotesize{td}}}}[k - k' \, ; \, k']  \, + \, n_{\mbox{\footnotesize{td}}}[k]
\end{eqnarray}}where $h_{_{\mbox{\footnotesize{td}}}}[n \, ; \, k]$ is the effective TD channel filter and the noise term
$n_{\mbox{\footnotesize{td}}}[k]$ is given by sampling at $t= k/B$ the continuous function
\begin{eqnarray*}
w_{rx}(t) \, \star \, n_{\mbox{\footnotesize{td}}}(t),
\end{eqnarray*}where $w_{rx}(t)$ is the matched filter at the receiver and $\star$ is linear convolution. Equation (\ref{paper2_eqn2}) can be expressed in a matrix form as follows. Arrange the transmitted information symbols as a $BT$ column vector
\begin{eqnarray}
\label{paper2_eqn3}
\left( {\bf x}_{\mbox{\footnotesize{td}}} \right)_{k+1} & \hspace{-3mm} = &  \hspace{-3mm} x_{\mbox{\footnotesize{td}}}[k] = x[k],
\end{eqnarray}
for $k=0,1,\cdots, BT-1$. Arrange the received TD samples as a $BT+K_1+K_2$ column vector 
\begin{eqnarray}
\label{paper2_eqn201}
\left( {\bf y}_{\mbox{\footnotesize{td}}} \right)_{k+1+K_1} & \hspace{-3mm} = & \hspace{-3mm}  y_{\mbox{\footnotesize{td}}}[k], 
\end{eqnarray}
for $k=-K_1,\cdots, BT-1+K_2$, where the constants $K_1, K_2 \in {\mathbb Z}$ are determined by the channel delay spread. Similarly, arrange the received sampled noise as a $BT+K_1+K_2$ vector
\begin{eqnarray}
\label{paper2_eqn197}
\left( {\bf n}_{\mbox{\footnotesize{td}}} \right)_{k+1+K_1} & \hspace{-3mm} = &  \hspace{-3mm} n_{\mbox{\footnotesize{td}}}[k],
\end{eqnarray}
for $k=-K_1,\cdots, BT-1+K_2$. Finally, arrange the effective channel filter as a $(BT + K_1 + K_2) \times BT$ matrix
\begin{eqnarray}
\label{paper2_eqn199}
\left( {\bf H}_{\mbox{\footnotesize{td}}} \right)_{k'+1,k+1}
& = & h_{_{\mbox{\footnotesize{td}}}}[k' - k - K_1 \, ; \, k].
\end{eqnarray}
Putting everything together, we obtain the matrix relation
\begin{eqnarray}
\label{paper2_eqn4}
{\bf y}_{\mbox{\footnotesize{td}}} & = & {\bf H}_{\mbox{\footnotesize{td}}} \, {\bf x}_{\mbox{\footnotesize{td}}} \, + \, {\bf n}_{\mbox{\footnotesize{td}}}.
\end{eqnarray}

\subsection{FDM}
\label{subsecfdmmatrixIO}
Direct calculation reveals that the {discrete} frequency-domain I/O relation induced from (\ref{paper2_eqn1}), is given by (see Equation ($25$) in Part I)  

{\vspace{-4mm}

\begin{eqnarray}
\label{paper2_eqn5}
y_{\mbox{\footnotesize{fd}}}[k] & = & \sum\limits_{k' \in {\mathbb Z}} x_{\mbox{\footnotesize{fd}}}[k'] \, h_{_{\mbox{\footnotesize{fd}}}}[k - k' \, ; \, k']  \, + \, n_{\mbox{\footnotesize{fd}}}[k]
\end{eqnarray}}
where $h_{_{\mbox{\footnotesize{fd}}}}[n \, ; \, k]$ is the effective FD channel filter and the noise term $n_{\mbox{\footnotesize{fd}}}[k]$ is obtained by sampling at $f=k/T$ the continuous function $w_{rx}(f) \star n_{\mbox{\footnotesize{fd}}}(f)$ where
\begin{eqnarray*}
 n_{\mbox{\footnotesize{fd}}}(f) & = & \int n_{\mbox{\footnotesize{td}}}(t) \, e^{-j 2 \pi f t} \, dt. 
\end{eqnarray*}
Arrange the transmitted information symbols as a $BT$ column vector
\begin{eqnarray}
\label{paper2_eqn6}
\left( {\bf x}_{\mbox{\footnotesize{fd}}} \right)_{k+1} & 
 \hspace{-3mm} = &  \hspace{-3mm}
{x}_{\mbox{\footnotesize{fd}}}[k] = x[k],
\end{eqnarray}
for $k=0,1,\cdots,BT-1$. Arrange the received samples as a $BT+L_1+L_2$ column vector
\begin{eqnarray}
\label{paper2_eqn210}
\left( {\bf y}_{\mbox{\footnotesize{fd}}} \right)_{k+1+L_1} & \hspace{-3mm} = & \hspace{-3mm} {y}_{\mbox{\footnotesize{fd}}}[k],
\end{eqnarray}
for $k=-L_1, \cdots, BT-1 + L_2$, where the constants $L_1, L_2 \in {\mathbb Z}$ are determined by the channel Doppler spread. Similarly, arrange the received noise samples as a $BT+L_1+L_2$ column vector
\begin{eqnarray}
\label{paper2_eqn205}
\left( {\bf n}_{\mbox{\footnotesize{fd}}} \right)_{k+1+L_1} & \hspace{-3mm} = &  \hspace{-3mm} n_{\mbox{\footnotesize{fd}}}[k],
\end{eqnarray}
for $k=-L_1,\cdots, BT-1+L_2$. Finally, arrange the FD channel filter as a $(BT + L_1 + L_2) \times BT$ matrix
\begin{eqnarray}
\label{paper2_eqn206}
\left( {\bf H}_{\mbox{\footnotesize{fd}}}  \right)_{k'+1,k+1} & \hspace{-3mm} = & \hspace{-3mm} h_{_{\mbox{\footnotesize{fd}}}}[k' - k - L_1\, ; \, k],
\end{eqnarray}
for $k'=0,1,\cdots,BT+L_1+L_2-1$ and $k=0,1,\cdots, BT-1$. Putting everything together, we obtain the matrix relation
\begin{eqnarray}
\label{paper2_eqn207}
{\bf y}_{\mbox{\footnotesize{fd}}} & = & {\bf H}_{\mbox{\footnotesize{fd}}} \, {\bf x}_{\mbox{\footnotesize{fd}}} \, + \, {\bf n}_{\mbox{\footnotesize{fd}}}.
\end{eqnarray}

\section{Impact of fading in the crystalline regime}
\label{secperfcsi}
In this Section, we compare uncoded BER performance of Zak-OTFS, TDM and FDM for the Veh-A channel model \cite{EVAITU}. Since we are focusing on the impact of fading, we suppose that the input-output (I/O) relation is perfectly known. We study the performance of Zak-OTFS as we move along the hyperbola $\tau_p  \cdot \nu_p = 1$ by choosing different Doppler periods $\nu_p$. We shall demonstrate that the performance of Zak-OTFS is superior in the crystalline regime, that performance approaches TDM as the Doppler period shrinks, and that performance approaches FDM as the delay period shrinks. 

Our Veh-A channel consists of six channel paths,
and the delay-Doppler spreading function is given by
\begin{eqnarray}
\label{paper2_eqn13}
h_{_{\mbox{\scriptsize{phy}}}}(\tau,\nu) & = & \sum\limits_{i=1}^6 h_i \, \delta(\tau - \tau_i) \, \delta(\nu - \nu_i).
\end{eqnarray}
where $h_i, \tau_i,$ and $\nu_i$ respectively
denote the gain, delay, and Doppler shift of the $i$-th channel path.
Table-\ref{tab1_paper2} lists the power-delay profile for the six channel paths. The maximum Doppler shift is $\nu_{max} = 815$ Hz, the Doppler spread is $1.63$ KHz, and the delay spread is $\tau_{max} \Define \max_i \tau_i \, - \, \min_i \tau_i = 2.5 \, \mu s$. 
The Doppler shift of the $i$-th path is modeled as $\nu_i = \nu_{max} \cos(\theta_i)$, where the variables $\theta_i, i=1,2,\cdots, 6$ are independent and distributed uniformly in the interval $[0 \,,\, 2\pi)$.
We fix the time duration $T$ of a data frame to be $T = 1.6$ ms, and we
fix the bandwidth $B$ to be $0.96$ MHz.

We now specify transmit and receive filters. For TDM,
the pulse shaping filter at the transmitter, and the matched
filter at the receiver, are both narrow TD sinc pulses
of bandwidth $B$ and TD pulse width proportional to $1/B$:
\begin{eqnarray}
w_{tx}(t) = w_{rx}(t) = \sqrt{B} \,sinc(B t).
\end{eqnarray}
For FDM, the transmit and receive filters are narrow FD sinc pulses with width proportional to $1/T$, so that the time-realization has duration $T$:
\begin{eqnarray}
w_{tx}(f) = w_{rx}(f) =\sqrt{T} \,  sinc(f T).
\end{eqnarray}
For Zak-OTFS, the DD domain transmit and receive filters are
the product of a narrow pulse in the delay domain with width proportional to $1/B$ and a narrow pulse in the Doppler domain with width proportional to $1/T$:
\begin{eqnarray}
w_{tx}(\tau,\nu) & \hspace{-3mm} = & \hspace{-3mm} w_{rx}(\tau, \nu)  =  \sqrt{B T} \, sinc(B \tau) \, sinc(T \nu).
\end{eqnarray}
We measure BER performance as a function of the received SNR,
which is the ratio of the power of the information carrying signal to the power of the AWGN in the received TD signal.
We normalize the complex channel gains by setting $\sum\limits_{i=1}^{6}{\mathbb E}[\vert h_i \vert^2] = 1$. We define the transmitted signal power $P_T$ to be the ratio of the average energy of the transmitted signal to the frame duration $T$:
\begin{eqnarray}
P_T & \Define & \frac{{\mathbb E}\left[ \int \vert {s}_{_{\mbox{\footnotesize{td}}}}(t) \vert^2 \, dt \right]}{T}.
\end{eqnarray}
If $N_0$ is the noise power spectral density of the AWGN ${n}_{_{\mbox{\footnotesize{td}}}}(t)$, then the noise power at the receiver is $N_0 B$, and
the signal-to-noise ratio (SNR) is given by
\begin{eqnarray}
\label{paper2_gamma_defeqn}
\gamma & \Define & \frac{P_T}{N_0 B}.
\end{eqnarray}
We assume that the Zak-OTFS receiver knows the effective
channel matrix ${\bf H}_{_{\mbox{\footnotesize{dd}}}}$
perfectly, that the TDM receiver knows ${\bf H}_{_{\mbox{\footnotesize{td}}}}$ perfectly, and that the
FDM receiver knows ${\bf H}_{_{\mbox{\footnotesize{fd}}}}$
perfectly. \emph{Note that perfect knowledge of the I/O relation does not imply perfect knowledge of the channel delay-Doppler spreading function.} 
The matrix-vector I/O relations given by 
(\ref{paper2_eqnyddhdd}), (\ref{paper2_eqn4}) and (\ref{paper2_eqn207}) have the same structure as that of a MIMO system. Hence, we use a
Linear Minimum Mean Squared Error (LMMSE) equalizer to detect the transmitted $4$-QAM information symbols.

\begin{table}
\caption{Power Delay Profile of Doubly-spread Veh-A Channel.}
\centering
\begin{tabular}{ | c || c |  c |  c | c |  c |  c |} 
  \hline
   Path no. $i$ & $1$ & $2$ & $3$ & $4$ & $5$ & $6$  \\
   \hline
  Rel. Delay $\tau_i$ ($\mu s$) & $0$ & $0.31$ & $0.71$ & $1.09$ & $1.73$ & $2.51$  \\ 
  \hline
  Rel. Power $\frac{{\mathbb E}[\vert h_i \vert^2]}{{\mathbb E}[\vert h_1 \vert^2]}$ (dB)      & $0$   &  $-1$  &  $-9$ &  $-10$ & $-15$  &  $-20$ \\
  \hline
\end{tabular}
\label{tab1_paper2}
\end{table}

\begin{figure}
\includegraphics[width=9cm, height=6.5cm]{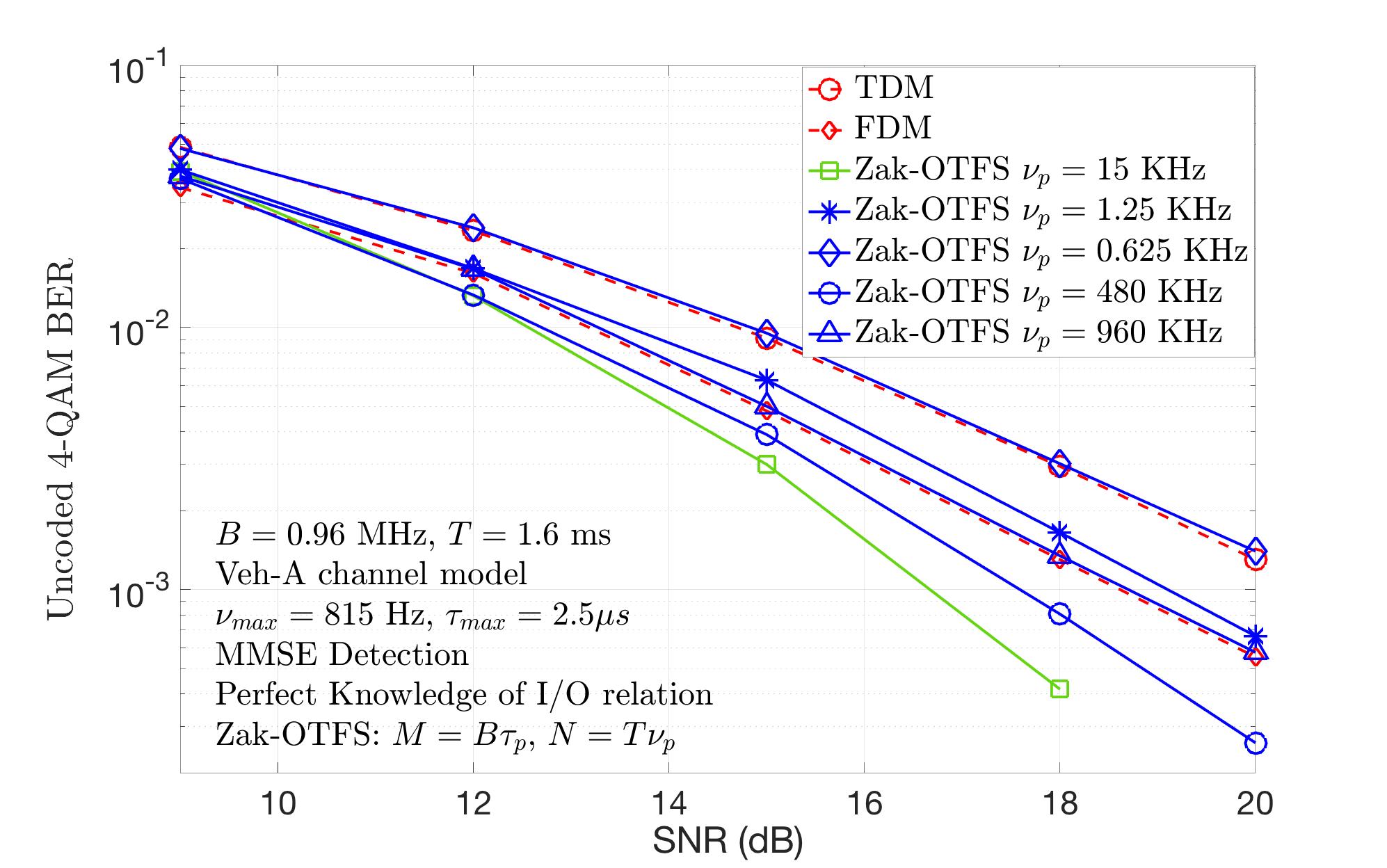}
\caption{BER performance of Zak-OTFS, TDM and FDM,
on a doubly-spread Veh-A channel as we traverse the hyperbola $\tau_p \cdot \nu_p = 1$. Performance of
Zak-OTFS is superior in the crystalline regime
($\nu_p = 15$ KHz), performance approaches TDM as the delay period grows ($\nu_p = 1.25$ and $0.625$ KHz) and performance approaches FDM as the Doppler period grows ($\nu_p = 480$ and $960$ KHz).}
\label{fig1}
\end{figure}
Fig.~\ref{fig1} illustrates how BER performance of Zak-OTFS
changes as we move along the hyperbola $\tau_p \cdot \nu_p = 1$, choosing Doppler periods $\nu_p = 0.625, 1.25, 15, 480$, and $960$ KHz. When $\nu_p = 15$ KHz, the crystallization
conditions hold, the I/O relation
is non-fading (see (\ref{paper2_nofading1})),
and BER performance is superior to both TDM and FDM. This is because the I/O relations for TDM and FDM exhibit fading on doubly spread channels.

As the Doppler period increases, first to $480$ KHz, then to $960$ KHz, the delay spread is no longer less than the delay period, and Zak-OTFS operates outside the crystalline regime.
Aliasing occurs along the delay axis, and the received symbol
energy varies (fades) along the Doppler axis.
When $\nu_p = 960$ KHz, the BER performance of Zak-OTFS
coincides with FDM, which is expected, given that FDM coincides
with Zak-OTFS in the limit of vanishing delay period.

As the Doppler period decreases, first to $1.25$ KHz, then to $0.625$ KHz, the Doppler spread is no longer less than the Doppler period, and Zak-OTFS operates outside the crystalline regime. Aliasing now occurs along the Doppler axis, and the received symbol energy varies (fades) along the delay axis.
When $\nu_p = 0.625$ KHz, the BER performance of Zak-OTFS almost coincides with TDM, which is expected, given that TDM
coincides with Zak-OTFS in the limit of vanishing Doppler period.

\begin{figure}[h]
\includegraphics[width=9.5cm, height=6.5cm]{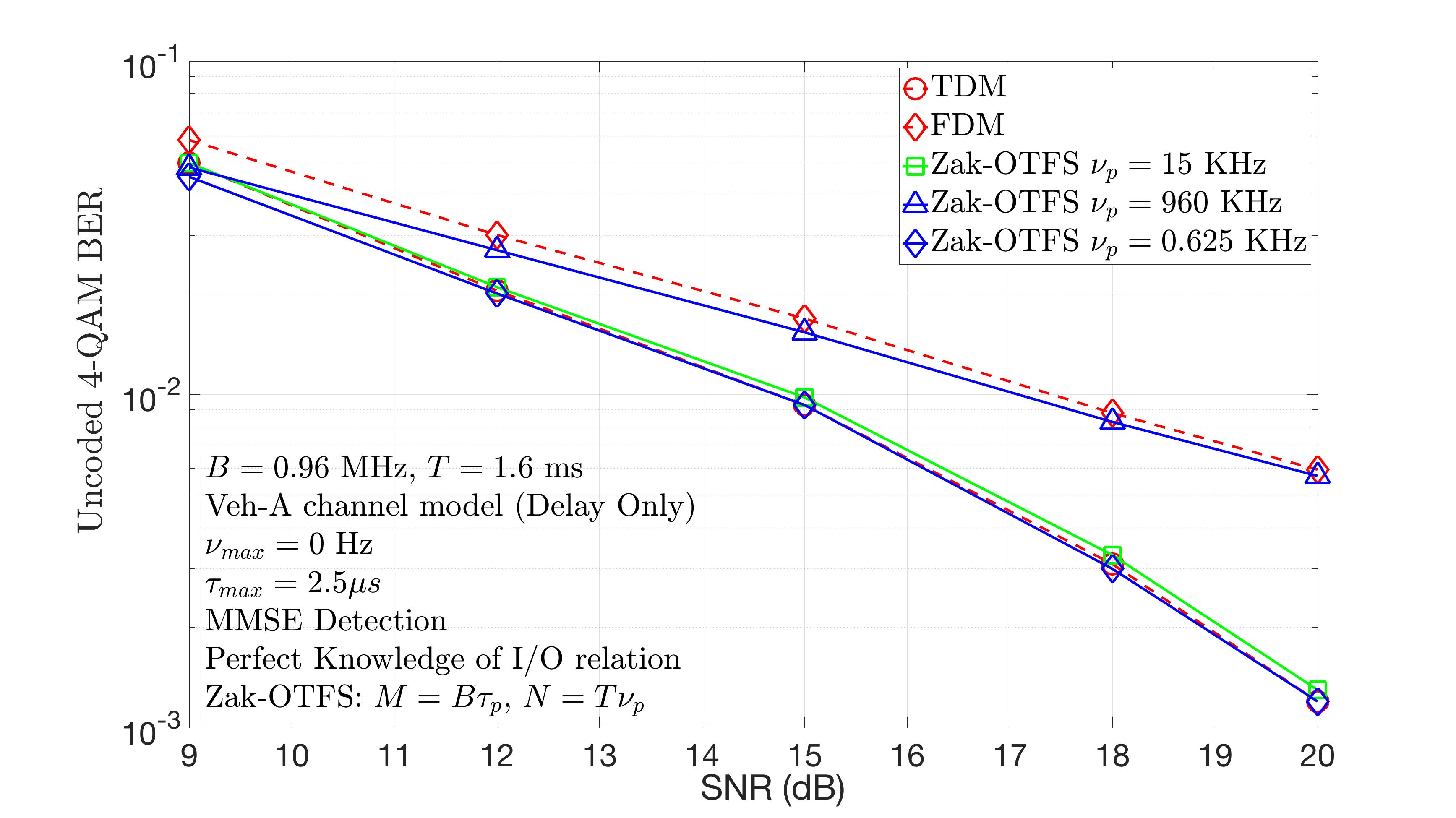}
\caption{BER performance of Zak-OTFS, TDM, and FDM on a delay-only Veh-A channel. The I/O relation for Zak-OTFS in the crystalline regime
($\nu_ p = 15$ KHz) is non-fading, hence BER performance matches that of TDM. Frequency selective fading degrades the BER performance of FDM. 
}
\label{fig2}
\end{figure}

A \emph{delay-only} Veh-A channel is matched to TDM and mismatched
to FDM. The channel is frequency-selective, not time-selective,
hence FDM exhibits fading, and TDM does not.
Fig.~\ref{fig2} compares BER performance of TDM and FDM with that of Zak-OTFS in a delay-only Veh-A channel (i.e., $\nu_i = 0, i=1,2,\cdots, 6$) for Doppler periods $\nu_p = 0.625, 15$, and $960$ KHz. When
$\nu_p = 0.625$ KHz, the delay period $\tau_p = 1.6$ ms is much larger
than the delay spread ($2.5 \, \mu s$), and BER performance is essentially the same as that of TDM. As long as we operate in the crystalline regime (for example, $\nu_p = 15$ KHz and $\tau_p = 66.66 \, \mu s$), there is no fading, and BER performance changes very little with change in $\nu_p$.
Outside the crystalline regime, when the Doppler period is large and the delay period is smaller than the delay spread (for example, $\nu_p = 960$ KHz and $\tau_p = 1.04 \, \mu s$), the BER performance degrades considerably. 

\begin{figure}[h]
\includegraphics[width=9.5cm, height=6.5cm]{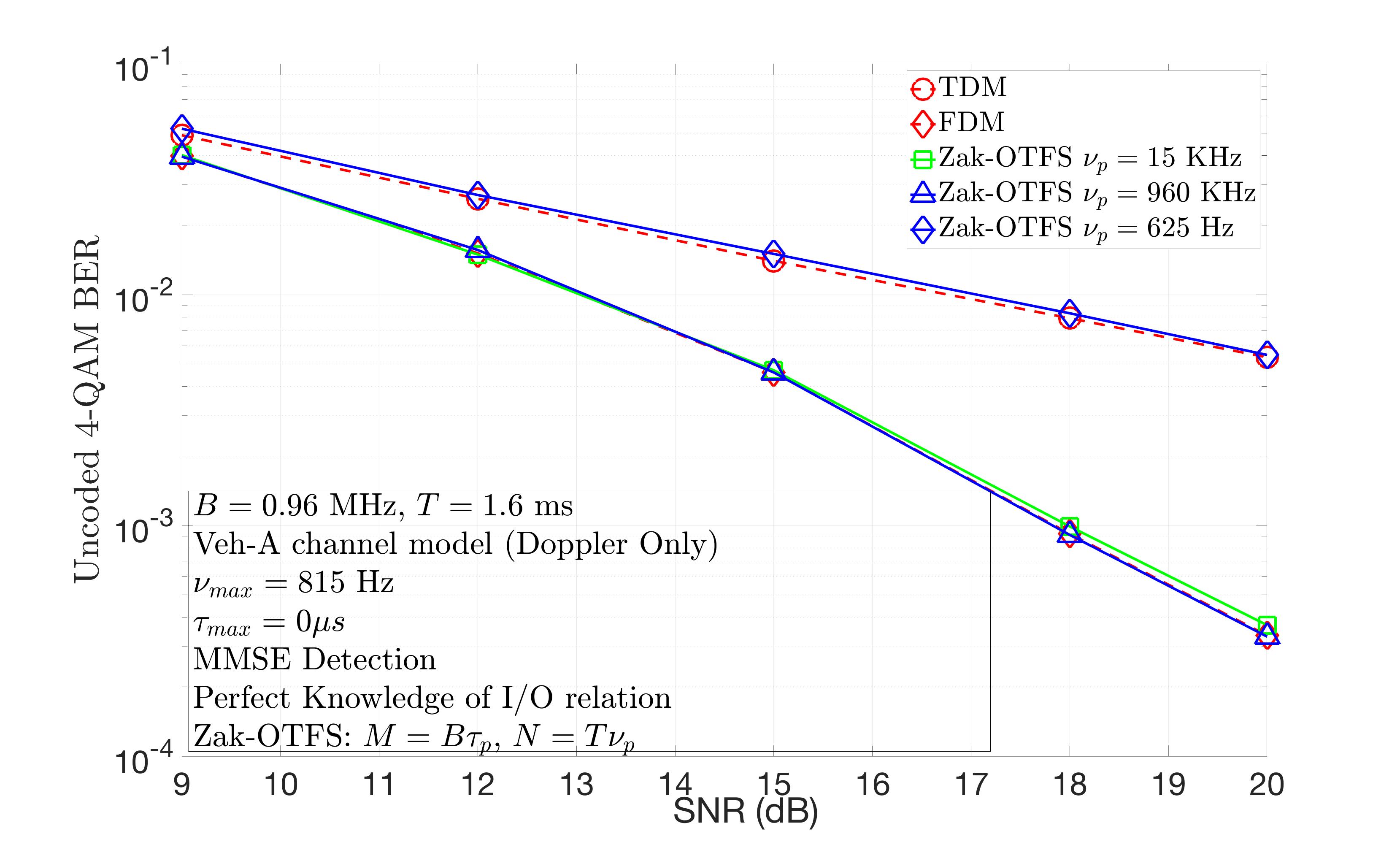}
\caption{BER performance of Zak-OTFS, TDM, and FDM,
on a Doppler-only Veh-A channel. 
The I/O relation for Zak-OTFS in the crystalline regime
($\nu_p = 15$ KHz) is non-fading,
hence BER performance matches that of FDM. Time
selective fading degrades BER performance of TDM.
}
\label{fig3}
\end{figure}

In contrast, a \emph{Doppler-only} Veh-A channel is matched
to FDM and mismatched to TDM. The channel is time-selective,
not frequency-selective, hence TDM exhibits fading and FDM does not. 
Fig.~\ref{fig3} compares BER performance of TDM and FDM with that
of Zak-OTFS in a Doppler-only Veh-A channel (i.e., $\tau_{max} = 0$), for Doppler periods $\nu_p = 0.625, 15$ and $960$ KHz.
When $\nu_p = 960$ KHz, the Doppler period is much larger than the Doppler spread ($1.63$ KHz), and BER performance is essentially the same as that of FDM. Again, when we operate in the crystalline regime (for example
$\nu_p = 15$ KHz and $\tau_p = 66.66 \, \mu s$) there is no fading, and the BER performance changes very little with change in $\nu_p$. Outside the crystalline regime
(for example $\nu_p = 0.625$ KHz), the BER performance degrades considerably.

\section{Model-free operation in the Crystalline regime}
\label{section5paper2}
In this section, we compare uncoded BER performance of Zak-OTFS, TDM and
FDM, in the typical scenario where the effective channel matrix is not
known perfectly at the receiver. The effective channel matrix
can be acquired from the I/O relation (see Section \ref{seciorelation}),
which in turn depends on the delay-Doppler spreading function
$h_{_{\mbox{\scriptsize{phy}}}}(\tau,\nu)$. The \emph{model-dependent} and \emph{model-free}
modes of operation correspond to different approaches to estimating
the effective channel matrix.

\textbf{Model-dependent operation:}
Here we impose a model on the delay-Doppler spreading function $h_{_{\mbox{\scriptsize{phy}}}}(\tau,\nu)$,
typically by prescribing a finite number of paths, and constraining their delay and Doppler shifts. Given this model, the receiver estimates
$h_{_{\mbox{\scriptsize{phy}}}}(\tau,\nu)$, then forms an estimate of the effective channel matrix. The accuracy of this estimate is limited
by time and bandwidth constraints on the pilot signal, and by any
mismatch between the channel model and the physical channel.

\textbf{Model-free operation:}
{Here the receiver estimates the effective DD domain channel filter
taps $h_{_{\mbox{\scriptsize{eff}}}}[k,l]$ directly, without reference to any model for the delay-Doppler spreading function $h_{_{\mbox{\scriptsize{phy}}}}(\tau,\nu)$.} {Although model-free operation can be chosen for both the crystalline and the non-crystalline regimes, communication reliability degrades in the non-crystalline regime
since the effective channel filter taps cannot be estimated accurately (see Section \ref{secNoncrystalline}). On the other hand, in the crystalline regime, the effective channel filter taps $h_{_{\mbox{\scriptsize{eff}}}}[k,l]$
can be estimated from the response to a single pilot symbol (see
Sections \ref{secCrystalline} and \ref{secCrysdecomp}).}

\underline{{Remark:}}
{The model-free approach considered here is a LTV channel generalization of the approach taken in LTI channels where only the taps of the effective discrete-time channel filter are estimated instead of estimating the continuous time impulse response of the underlying physical channel.}

The model-dependent mode of operation is a challenge for the Veh-A channel introduced in Section \ref{secperfcsi} (with sinc pulse shaping filters).
Since channel bandwidth $B = 0.96$ MHz, the delay domain resolution is $1/B \approx 1.04 \, \mu s$, and the first three paths introduce delay shifts in the interval $[ 0 \,,\, 0.71] \, \mu s$. These paths are not separable, and so cannot be estimated accurately.

However, model-free operation is still possible for TDM, FDM, and for Zak-OTFS in the crystalline regime ($\nu_p = 15$ KHz). We estimate the effective channel filter taps from the response to a pilot frame with a single high-energy pilot and no information symbols. We want to start
from ground truth, so we suppose that the received pilot is not subject to AWGN. For Zak-OTFS we locate the single pilot at the center $(M/2 \,,\, N/2)$ of the fundamental period, and we use (\ref{paper2_predicteqn1}) to estimate the effective filter taps. For TDM and FDM we locate the single pilot
at $k = BT/2$, estimate the effective channel filter taps
$h_{_{\mbox{\footnotesize{td/fd}}}}[n \, ; \, k=BT/2], n \in {\mathbb Z}$,
and simply reuse this estimate for all other taps $h_{_{\mbox{\footnotesize{td/fd}}}}[n \, ; \, k]$ $n \in {\mathbb Z}, k=0,1,\cdots, BT-1, k \ne BT/2$.

We attempt model-dependent operation, using the pilots described above to
estimate the complex channel gain, delay and Doppler shift of each channel
path. We interpret DD points with significant energy as the locations
of channel path delays and Doppler shifts. Given the locations of delay and
Doppler shifts, the received pilot signal depends linearly on the channel path gains, so we use least squares to estimate the vector of complex channel gains, then reconstruct ${\bf H}_{_{\mbox{\footnotesize{dd}}}}$
using (\ref{paper2_eqn12}). 

\begin{figure}[h]
\includegraphics[width=9cm, height=6.5cm]{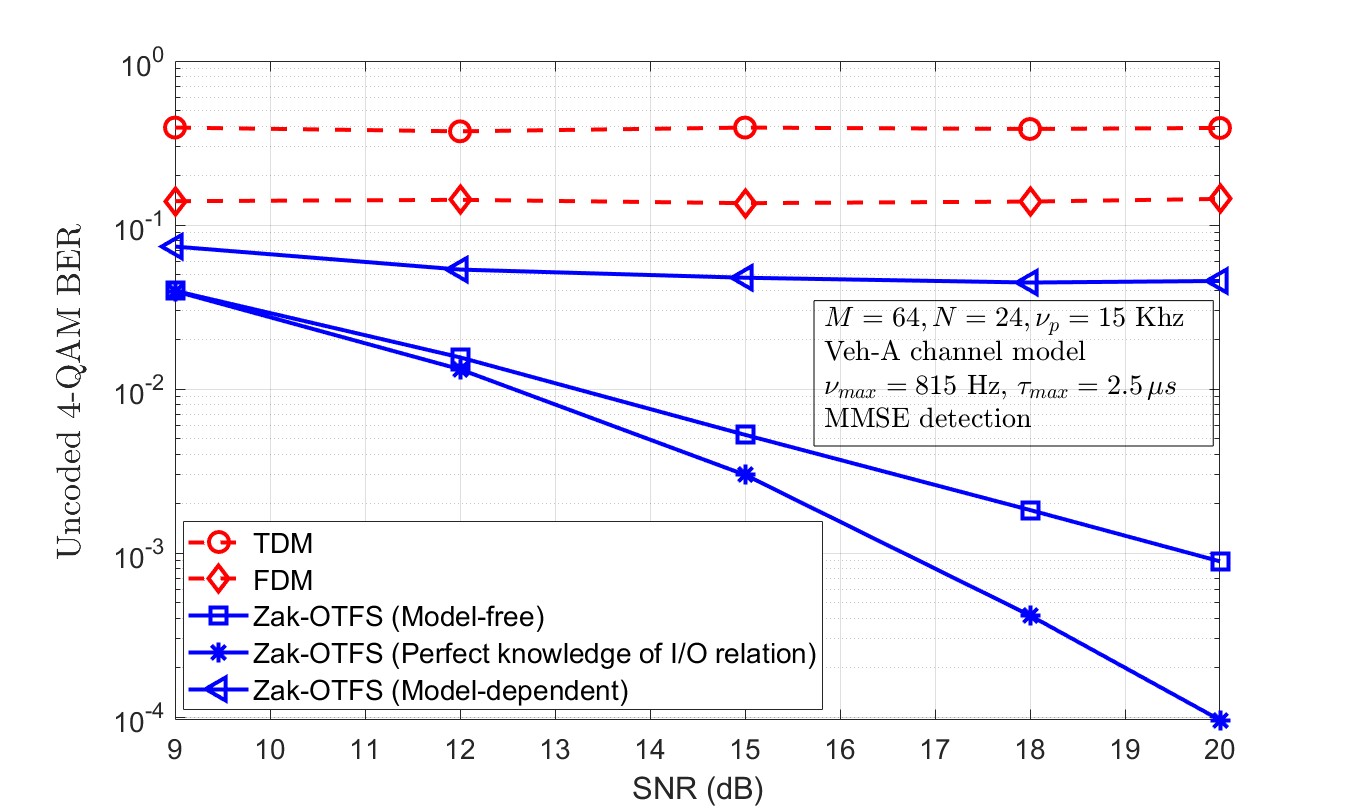}
\caption{BER performance of Zak-OTFS, TDM, and FDM on the Veh-A channel (Section IV), where signal bandwidth and duration is insufficient to estimate channel path delays and Doppler shifts. In the crystalline regime ($\nu_p = 15$ KHz), BER performance of the model-free mode of Zak-OTFS is only slightly worse than performance with perfect knowledge of the I/O relation.}
\label{fig4}
\end{figure}

Fig.~\ref{fig4} compares BER performance of the model-dependent and model-free modes of Zak-OTFS in the crystalline regime. It also includes BER performance of TDM and FDM, which is poor, and which does not improve with increasing SNR. Since the channel is doubly spread, the I/O relation for TDM/FDM is non-stationary in the TD/FD, and it cannot be accurately predicted from the response to a single pilot at $k = BT/2$. Hence, the estimate of the effective channel matrix is inaccurate, and the probability of mis-detection is high. BER performance of the model-dependent mode of Zak-OTFS is slightly better, but it also exhibits a high error floor because insufficient frame bandwidth and duration precludes accurate estimation of channel path gains, delays and Doppler shifts.\footnote{\footnotesize{For example, for the Veh-A channel, the first three paths
introduce delay shifts which lie in the interval $[0 \,,\, 0.71 ] \, \mu s$.
With a channel bandwidth of $B = 0.96$ MHz, the delay domain resolution is $1/B  \approx 1.04 \, \mu s$ which is more than $0.71 \, \mu s$ and therefore
the first three dominant paths are not separable/resolvable.
These non-separable paths cannot be estimated accurately.}} BER performance of the model-free mode of Zak-OTFS is considerably better, only slightly worse than performance with perfect knowledge of the I/O relation.

\begin{figure}[h]
\hspace{-2mm}
\includegraphics[width=9cm, height=6.5cm]{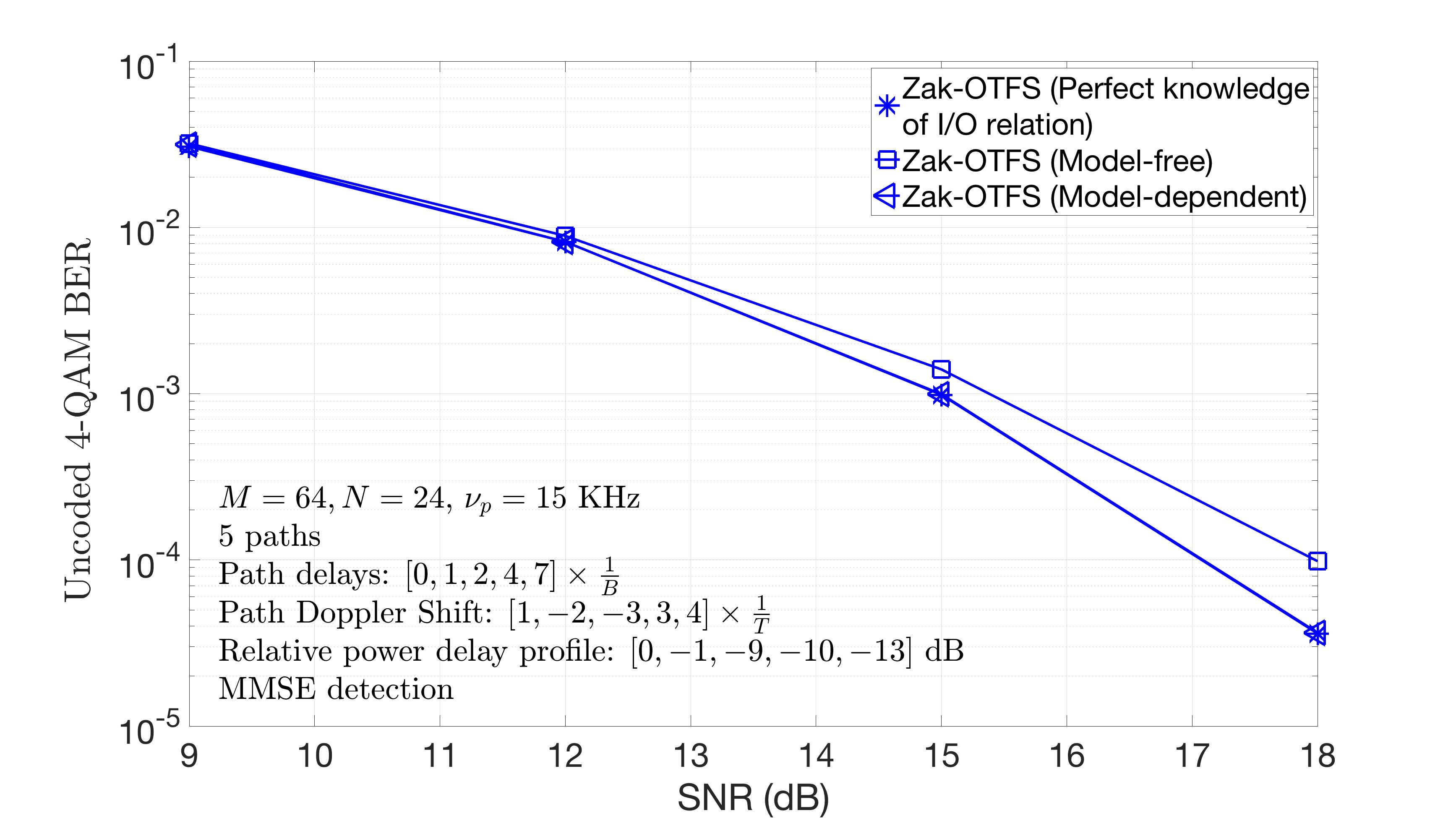}
\caption{
BER performance of model-dependent and model-free Zak-OTFS on a $5$-path resolvable channel, where the delay shifts are integer multiples of the delay resolution $1/B$, and the Doppler shifts are integer multiples of the Doppler resolution $1/T$. Model-dependent performance coincides with that of performance with perfect knowledge of the I/O relation. Model-free performance is only slightly inferior.
}
\label{fig5}
\end{figure}

Next, through Fig.~\ref{fig5} we explore what we lose when it is possible to learn the channel, but we choose to operate model-free. Again, we fix the time duration $T$ of a data frame to $1.6$ ms, and we fix the bandwidth $B$ to $0.96$ MHz. There are $M=64$ delay bins and $N=24$ Doppler bins. Again, the delay spread is roughly $2.5 \, \mu s$, and the Doppler spread is $1.63$ KHz. We consider a $5$-path resolvable channel, where the delay shifts are integer multiples of the delay resolution $1/B$, and the Doppler shifts are integer multiples of the Doppler resolution $1/T$. To be precise, the path delays are $[0, 1, 2, 4, 7] \times 1/B$, the Doppler shifts are $[1, -2, -3, 3, 4]  \times 1/T$, and the relative average power of the paths is $[0, -1, -9, -10, -13]$ dB. Again, we consider Zak-OTFS operating in the crystalline regime ($\nu_p = 15$ KHz). 

Fig.~\ref{fig5} illustrates that for this resolvable channel, model-dependent performance coincides with that of performance with perfect knowledge of the I/O relation. This is expected since estimation of the channel spreading function is accurate when paths are resolvable. Further, model-free performance is only slightly inferior, despite the high Doppler spread of $4.375$ KHz. Why the small degradation? Pulse shaping filters cause self-interaction/aliasing between the received pilot in the fundamental period and its quasi-periodic replicas. Though this aliasing is small in the crystalline regime, it is non-zero.\footnote{\footnotesize{In the Zak-OTFS I/O relation, the output DD signal is a twisted convolution of the effective discrete DD channel filter $h_{_{\mbox{\scriptsize{eff}}}}[k,l]$ with the input information signal $x_{_{\mbox{\footnotesize{dd}}}}[k,l]$, and therefore each information symbol is spread in the DD domain by an amount equal to the DD spread/width of $h_{_{\mbox{\scriptsize{eff}}}}[k,l]$. Since $h_{_{\mbox{\scriptsize{eff}}}}[k,l]$ is sampled from the effective continuous DD channel filter $h_{_{\mbox{\scriptsize{eff}}}}(\tau,\nu)$ at integer multiples of $1/B$ and $1/T$ along the delay and Doppler domains respectively (see equation ($36$) in Part I), the DD spread/width of $h_{_{\mbox{\scriptsize{eff}}}}[k,l]$ is directly related to the DD spread/width of $h_{_{\mbox{\scriptsize{eff}}}}(\tau,\nu)$. Further, $h_{_{\mbox{\scriptsize{eff}}}}(\tau,\nu) = w_{rx}(\tau,\nu) *_{\sigma} h_{_{\mbox{\scriptsize{phy}}}}(\tau,\nu)  *_{\sigma} w_{tx}(\tau,\nu)$ is a twisted convolution of the channel DD spreading function and the transmit and receive pulse shaping filters (see equation ($36$) in Part I). The DD domain spread of $h_{_{\mbox{\scriptsize{eff}}}}(\tau,\nu)$ is therefore the sum of the DD spread/width of  $h_{_{\mbox{\scriptsize{phy}}}}(\tau,\nu)$ and the DD spread/width of the transmit and receive DD pulse shaping filters $w_{tx}(\tau,\nu)$ and $w_{rx}(\tau,\nu)$. Hence the maximum effective DD spread of $h_{_{\mbox{\scriptsize{eff}}}}(\tau,\nu)$ is more than $\tau_{max}$ along the delay domain and more than $2 \nu_{max}$ along the Doppler domain.}}

\begin{figure}[h]
\hspace{-7mm}
\includegraphics[width=9.4cm, height=6.5cm]{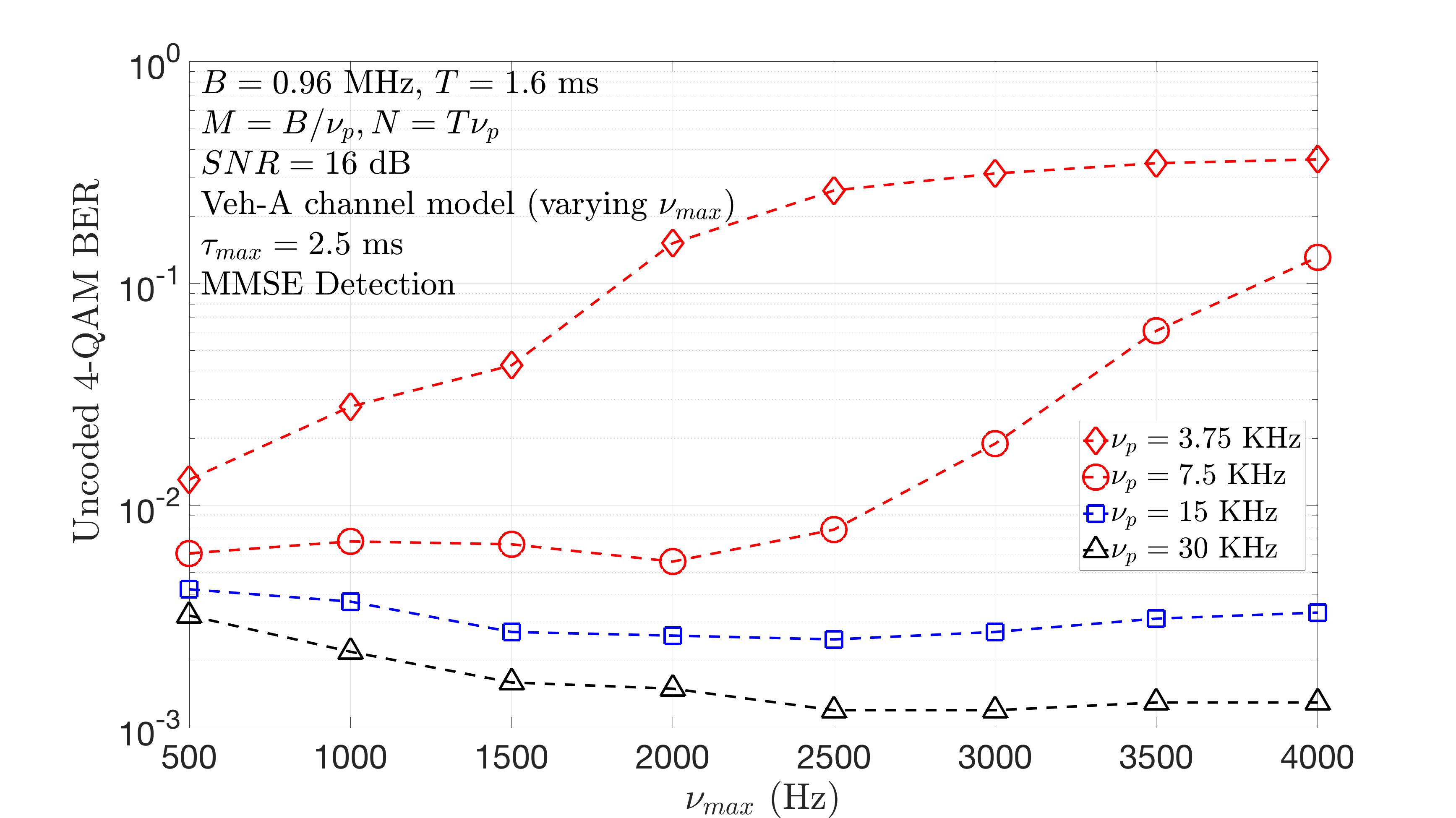}
\caption{. Model-free performance as a function of $\nu_{max}$ for the Veh-A channel introduced in Section \ref{secperfcsi}. When the Doppler spread $2 \nu_{max}$ is significantly less than $\nu_p$, performance does not degrade as $\nu_{max}$ increases. When the Doppler spread $2 \nu_{max}$ is close to $\nu_p$, performance degrades because of Doppler domain aliasing. When operating deep in the crystalline regime, BER performance is consistently excellent over a wide range of Dopplers.}
\label{fig6}
\end{figure}

Next, through Fig.~\ref{fig6} we explore model-free performance in the crystalline regime as we traverse the hyperbola $\tau_p \cdot \nu_p = 1$, moving towards FDM. We consider the Veh-A channel introduced in Section \ref{secperfcsi}, fixing the SNR at $16$ dB. We measure performance for Doppler periods $\nu_p = 3.75, 7.5, 15$, and $30$ KHz, as $\nu_{max}$ varies between $500$ Hz and $4$ KHz. For $\nu_p = 15, 30$ KHz, we are deep in the crystalline regime, since the Doppler period is significantly larger than the Doppler spread (which is at most $8$ KHz), and the delay period ($66.66, 33.33 \, \mu s$) is significantly larger than the delay spread ($2.5 \, \mu s$). The spacing between quasi-periodic replicas limits self-interaction, even with sinc pulse shaping filters. The Zak-OTFS I/O relation is predictable, and we are able to accurately estimate the taps of the effective channel filter. Fig.~\ref{fig6} illustrates that increasing the spacing from $15$ KHz to $30$ KHz improves performance slightly. For both Doppler periods, the BER performance is excellent, and almost invariant to increasing Doppler spread. 

BER performance changes as we reduce the Doppler period to $7.5$ KHz. When the Doppler spread $2 \nu_{max}$ is less than $4$ KHz, the interaction between the received DD pilot in the fundamental period and its quasi-periodic replicas is not significant. As the Doppler spread increases beyond $4$ KHz, BER increases steadily. When the Doppler spread is the same as the Doppler period ($\nu_{max} = 3.75$ KHz), BER performance degrades completely. This is because the sinc pulse shaping filters leak energy outside their ideal delay width ($1/B = 1.04 \, \mu s$) and Doppler width ($1/T = 625$ Hz). When we reduce the Doppler period to $3.75$ KHz, the BER performance starts to degrade earlier at $\nu_{max} = 500$ Hz.

\begin{figure}[h]
\hspace{-3mm}
\includegraphics[width=9.4cm, height=6.5cm]{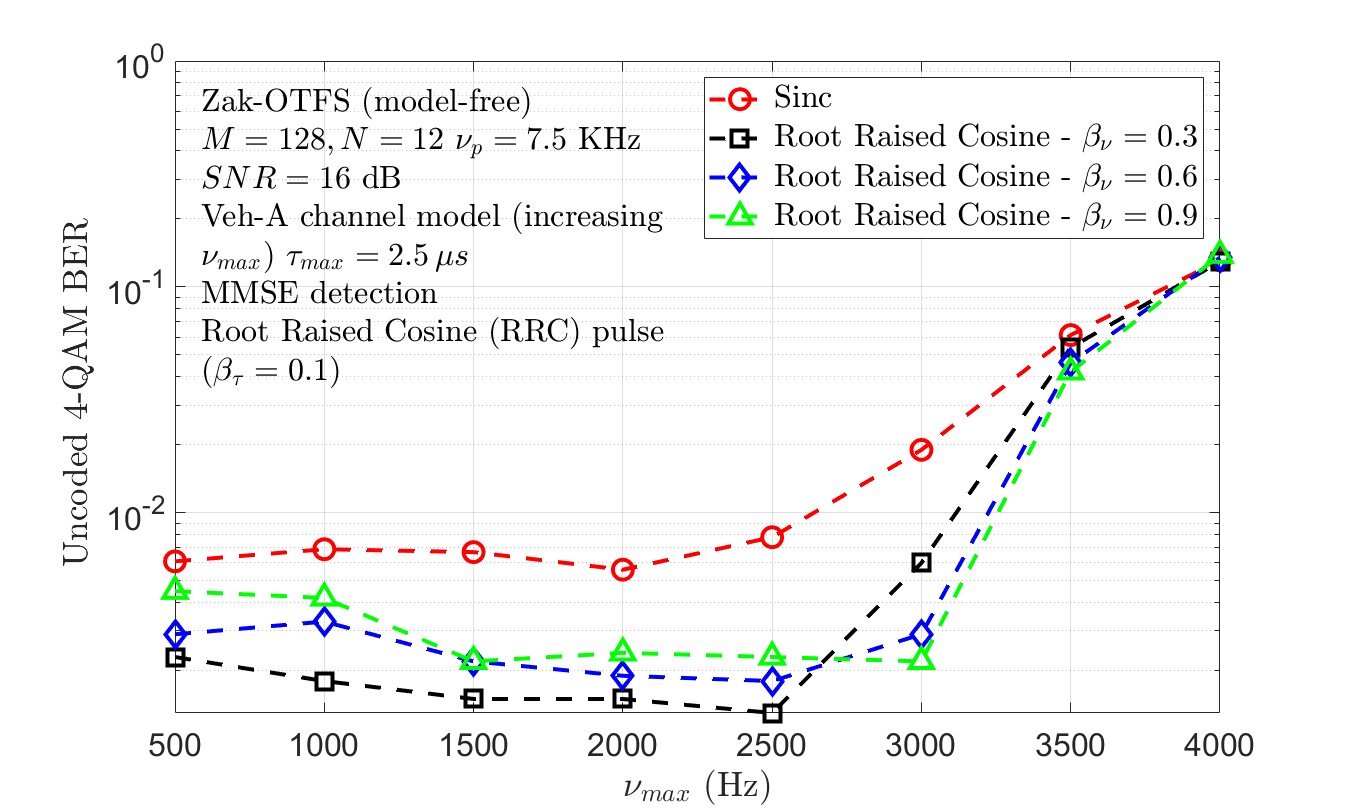}
\caption{Better pulse shaping filters extend the region of reliable model-free operation. BER performance, for the sinc pulse, and for RRC pulses with $\beta_{\tau} = 0.1$, and $\beta_{\nu} = 0.3, 0.6$, and $0.9$, as $\nu_{max}$ varies between $500$ Hz and $4$ KHz.
}
\label{fig7}
\end{figure}

Next, through Fig.~\ref{fig7} we show that better pulse shaping filters extend the region of reliable model-free operation by reducing aliasing. We consider DD domain Root Raised Cosine (RRC) pulses (specified in (\ref{rrcpulse_eqn1})), parameterized by a roll-off parameter $\beta_{\tau}$ that controls localization in delay, and a parameter $\beta_{\nu}$ that controls localization in Doppler. The sinc pulse corresponds to $\beta_{\tau} = \beta_{\nu} = 0$, and localization improves as the parameters $\beta_{\tau}, \beta_{\nu}$, increase from $0$ to $1$. Better localization implies lesser DD domain aliasing
and therefore accurate prediction/estimation of the I/O relation for a higher Doppler spread). We fix $\beta_{\tau} = 0.1$, and for $\beta_{\nu} = 0.3, 0.6$, and $0.9$, we measure BER performance over the Veh-A channel as $\nu_{max}$ varies between $500$ Hz and $4$ KHz (fixed SNR of $16$ dB, $M = 128$, $N = 12$, $\nu_p = 7.5$ KHz). Fig.~\ref{fig7} illustrates that increasing $\beta_{\nu}$ from $0$ (a sinc pulse) to $0.3$ extends the range of Doppler spreads for which BER performance is flat (from $4$ KHz to $5$ KHz). Increasing $\beta_{\nu}$ further, to $0.6, 0.9$, extends the range still further. The cost of introducing better filters is a reduction in spectral efficiency. When we replace a sinc filter by a RRC filter, we increase the frame duration by a factor $(1 + \beta_{\nu})$, and we increase the frame bandwidth by $(1 + \beta_{\tau})$.

\begin{figure*}
\centering
\includegraphics[width=17cm, height=5.4cm]{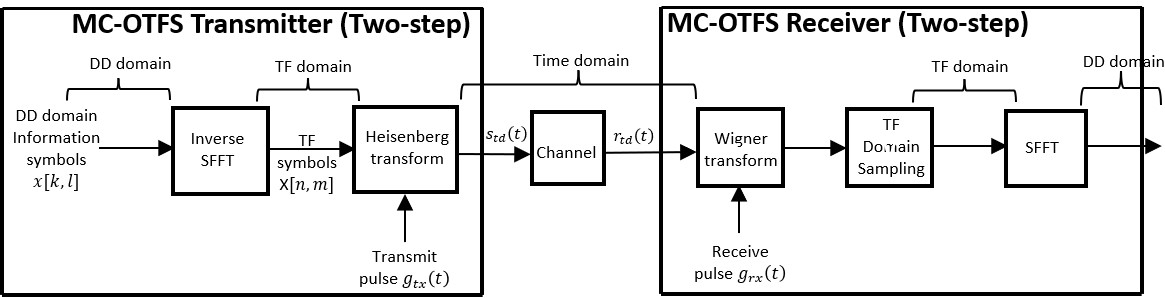}
\caption{Signal processing for MC-OTFS (two step).}
\label{figmcotfstwosteppaper2}
\end{figure*}
\begin{figure*}
\centering
\includegraphics[width=17cm, height=5.4cm]{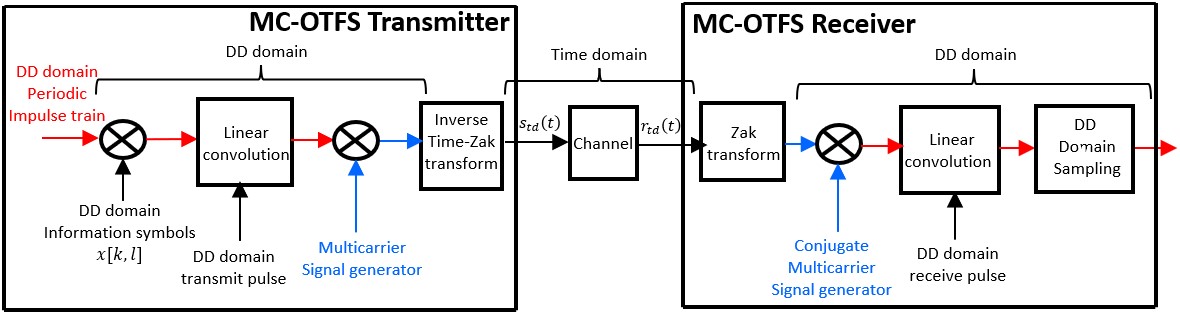}
\caption{Zak transform based equivalent signal processing
for the two-step MC-OTFS.}
\label{figmcotfspaper2}
\end{figure*}

\section{Multicarrier Approximations to Zak-OTFS}
\label{secotfsvariants}
So far, all existing work on OTFS presumes a
two-step modulation where the DD domain information symbols are first transformed to the time-frequency (TF) domain {(using the Inverse Symplectic Finite Fourier transform or Inverse SFFT \cite{RH1})}. The resulting TF symbols are then converted to a TD transmit signal using the Heisenberg transform (which is essentially an OFDM modulator) \cite{RH1} (see Fig.~\ref{figmcotfstwosteppaper2}). The compatibility of this two-step modulation with existing 4G/5G modems is one of the primary reasons why it was proposed.

This two-step modulation is fundamentally different from Zak-OTFS. One can think of it as a multicarrier approximation to Zak-OTFS, which we refer to here as MC-OTFS.
Interestingly, the two-step implementation of MC-OTFS can be {cast} in the framework of the Zak-transform as shown in Fig.~\ref{figmcotfspaper2}.

The simplicity of Zak-OTFS signal processing (shown in Fig.~\ref{figzakotfspaper2}) results from the fact that channels and filters both act by twisted convolution and all signals in the signal processing chain are quasi-periodic. Since twisted convolution is associative, the I/O relation admits a simple structure: the output signal is the twisted convolution of the input signal with an effective DD channel filter (see (\ref{paper2_eqn229})). This particular structure becomes predictable in the crystalline regime, implying that, in the crystalline regime, the complete I/O relation can be accurately estimated from the channel response to a single  DD pilot. In contrast, as we will see now, the I/O relation of MC OTFS cannot be expressed as a simple twisted convolution and, consequently, it does not appear to satisfy any reasonable form of predictability. That is, it is not clear how to estimate the I/O  relation from the observed channel response to a single DD pilot. 

A careful study of the Zak-transform implementation of MC-OTFS reveals that its underlying signal processing is fundamentally different and more complicated than that of Zak-OTFS. While the signal processing of Zak-OTFS comprises of pure cascade of twisted convolutions, the signal processing of MC-OTFS comprises of a mixture of linear convolution, multiplication by a quasi-periodic function and twisted convolution. As shown in Fig.~\ref{figmcotfspaper2}, in MC-OTFS, the information is encoded as a discrete periodic function (instead of discrete quasi-periodic function as in Zak-OTFS) on the DD domain, given by
\begin{eqnarray}
    x(\tau,\nu) & \hspace{-3mm} = & \hspace{-3mm} \sum\limits_{k, l \in {\mathbb Z}} x[k,l] \, \delta\left( \tau - \frac{k \tau_p}{M} \right) \, \delta\left( \nu - \frac{l \nu_p}{N} \right).
\end{eqnarray}
Next step is linear convolution (instead of twisted convolution as in Zak-OTFS) of the information with a transmit filter  $w_{tx}(\tau, \nu)$ resulting in a periodic continuous DD domain function. The transmit pulse is usually taken to be the SFFT of a TF window (whose time and frequency support are the time duration and bandwidth of the MC-OTFS frame). Next step is to convert the periodic DD domain function to a quasi-periodic one. This is achieved by multiplication with a fixed quasi-periodic signal $G_{_{\mbox{\footnotesize{dd}}}}(\tau, \nu)$ called the multi-carrier generator, which is nothing but the Zak transform of the TD transmit pulse $g_{tx}(t)$ in the two-step OTFS modulator. The transmitted signal is the inverse Zak transform of the resulting quasi-periodic signal. At the receiver, the Zak transform of the received signal which is quasi-periodic is converted into a periodic continuous function by means of multiplication with the complex conjugate of the multi-carrier signal generator. This is then followed by linear convolution with a receive filter $w_{rx}(\tau, \nu)$ resulting in $y^{w_{rx}}(\tau, \nu)$ which is then sampled. Ignoring AWGN effect, the relation between the information function $x(\tau, \nu)$ and the received filtered function $y^{w_{rx}}(\tau, \nu)$ is given in Table-\ref{tabeqn} (here $\star$ denotes linear convolution and $\cdot$ denotes multiplication). The counter part relation for Zak-OTFS is also given in this table. The main observation is that in Zak-OTFS, both the channel and the filters act through twisted convolution, hence due to associativity, the end-to-end signal processing is equivalent to a single twisted convolution with the effective channel filter $h_{_{\mbox{\scriptsize{eff}}}}(\tau, \nu)$. In contrast, since the MC-OTFS I/O relation is a mix of linear
convolution, multiplication and twisted convolution, it cannot be expressed as a simple action with some effective filter.

\begin{table*}
\vspace{-2mm}
\caption{I/O relation for Zak-OTFS and MC-OTFS}
\centering
\begin{tabular}{ | c | } 
  \hline
    MC-OTFS I/O relation   \\
  \hline
   \\
  $y^{w_{rx}}(\tau, \nu) \,  = \, w_{rx}(\tau, \nu) \, \star  \, \left[G_{_{\mbox{\footnotesize{dd}}}}^*(\tau, \nu) \cdot \left( h_{_{\mbox{\scriptsize{phy}}}}(\tau,\nu) *_{\sigma} \left\{ G_{_{\mbox{\footnotesize{dd}}}}(\tau, \nu) \cdot \left[ w_{tx}(\tau, \nu) \star x(\tau,\nu) \right]  \,  \right\} \right) \right]$ \\
   \\
  \hline
   Zak-OTFS I/O relation   \\
  \hline
   \\
  $y^{w_{rx}}_{_{\mbox{\footnotesize{dd}}}}(\tau, \nu)  \,  =  \,  w_{rx}(\tau, \nu) \, *_{\sigma}  \, h_{_{\mbox{\scriptsize{phy}}}}(\tau,\nu) \, *_{\sigma} \,   w_{tx}(\tau, \nu)  \, *_{\sigma} \, x_{_{\mbox{\footnotesize{dd}}}}(\tau,\nu)  \, = \,  h_{_{\mbox{\scriptsize{eff}}}}(\tau,\nu) \, *_{\sigma} \, x_{_{\mbox{\footnotesize{dd}}}}(\tau,\nu)$  \\
  \hline
\end{tabular}
\label{tabeqn}
\end{table*}

It is illuminating to observe, how MC-OTFS and Zak-OTFS evolve as we move towards TDM by shrinking the Doppler period $\nu_p$. As we traverse the hyperbola $\tau_p \cdot \nu_p = 1 $ along the limit $\nu_p \rightarrow 0$, the Zak transform converges to identity and Zak-OTFS converges to TDM whose I/O relation for delay-only channels is given by linear convolution with an effective channel filter, hence it is predictable. In contrast, as $\nu_p \rightarrow 0$, MC-OTFS does not converge to TDM. The limit in this case is a non-intuitive modulation whose I/O relation cannot be expressed as a simple linear convolution of an input signal with an effective TD channel filter.
\begin{figure}
\hspace{-6mm}
\includegraphics[width=9.0cm, height=5.5cm]{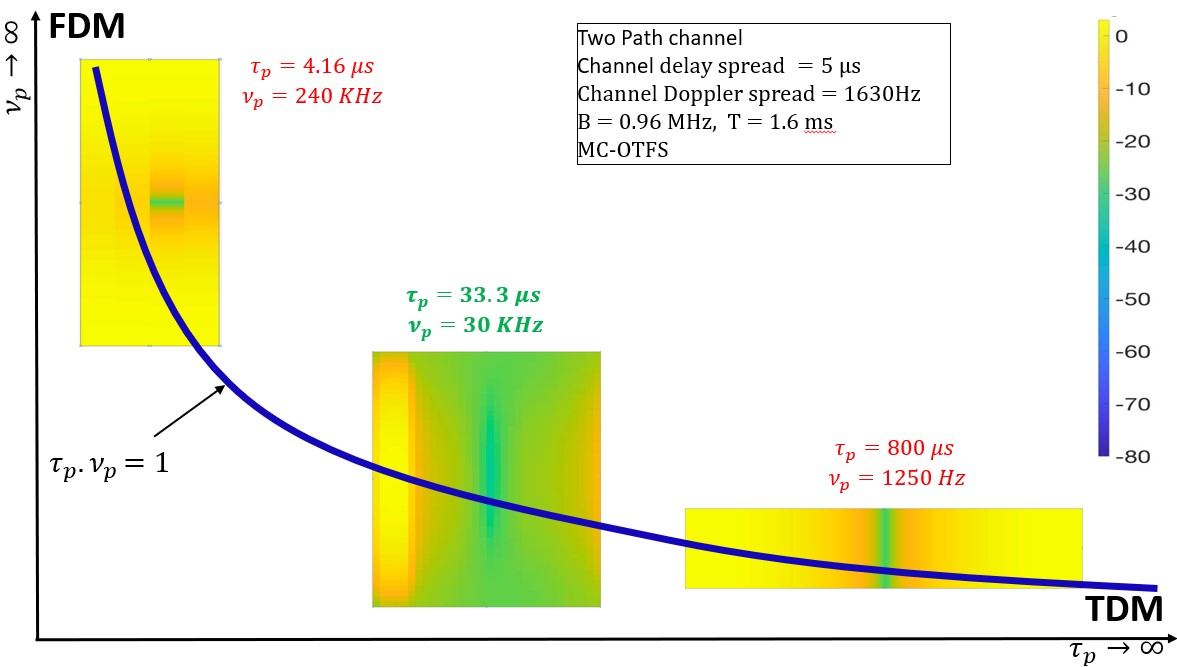}
\caption{Heatmap showing Relative Prediction Error (RPE), in dB, for MC-OTFS (with sinc pulses), as a function of delay (horizontal axis), and Doppler (vertical axis). RPE for MC-OTFS is larger than that for Zak-OTFS.}
\label{fig3paper2_mcotfs}
\end{figure}

Recall that Fig.~\ref{fig3paper2} depicts the relative prediction error for Zak-OTFS as a two-dimensional heatmap, for three different points on the period hyperbola $\tau_p \cdot \nu_p = 1$. Fig.~\ref{fig3paper2_mcotfs} depicts the relative prediction error for MC-OTFS, with respect to the same two-path channel, and the same transmit and receive filters (given by sinc pulses). The multi-carrier generator is taken to be the Zak transform of a rectangular TD pulse $g(t)$ of duration $\tau_p$. The MC-OTFS I/O relation is derived from the continuous I/O relation given in Table-\ref{tabeqn}.
In this relation the input is a discrete periodic function obtained by extending $x[k,l]$ periodically with periods $M = B \tau_p$ along delay and $N = T \nu_p$ along Doppler respectively. The output is obtained by sampling $y^{w_{rx}}(\tau, \nu)$
at integer multiples of $1/B$ along the delay and $1/T$ along Doppler. As for Zak-OTFS, we estimate the MC-OTFS I/O relation from the received response to a pilot impulse located at $(M/2, N/2)$.
The point $\nu_p = 1.25$ KHz, $\tau_p = 800 \, \mu s$ represents the delay asymptote outside the crystalline regime, and the point $\nu_p = 240$ KHz, $\tau_p = 4.16 \, \mu s$ represents the Doppler asymptote outside the crystalline regime. For these points, the relative prediction error is high for both Zak-OTFS and MC-OTFS. The point $\nu_p = 30$ KHz, $\tau_p = 33.3 \, \mu s$ represents the crystalline regime where relative prediction error for MC-OTFS is larger than for Zak-OTFS (the heatmap in Fig.~\ref{fig3paper2_mcotfs} is a mixture of yellow and green, whereas the heatmap in Fig.~\ref{fig3paper2} is mostly green). 
\begin{figure}[h]
\hspace{-7mm}
\includegraphics[width=9.9cm, height=6.5cm]{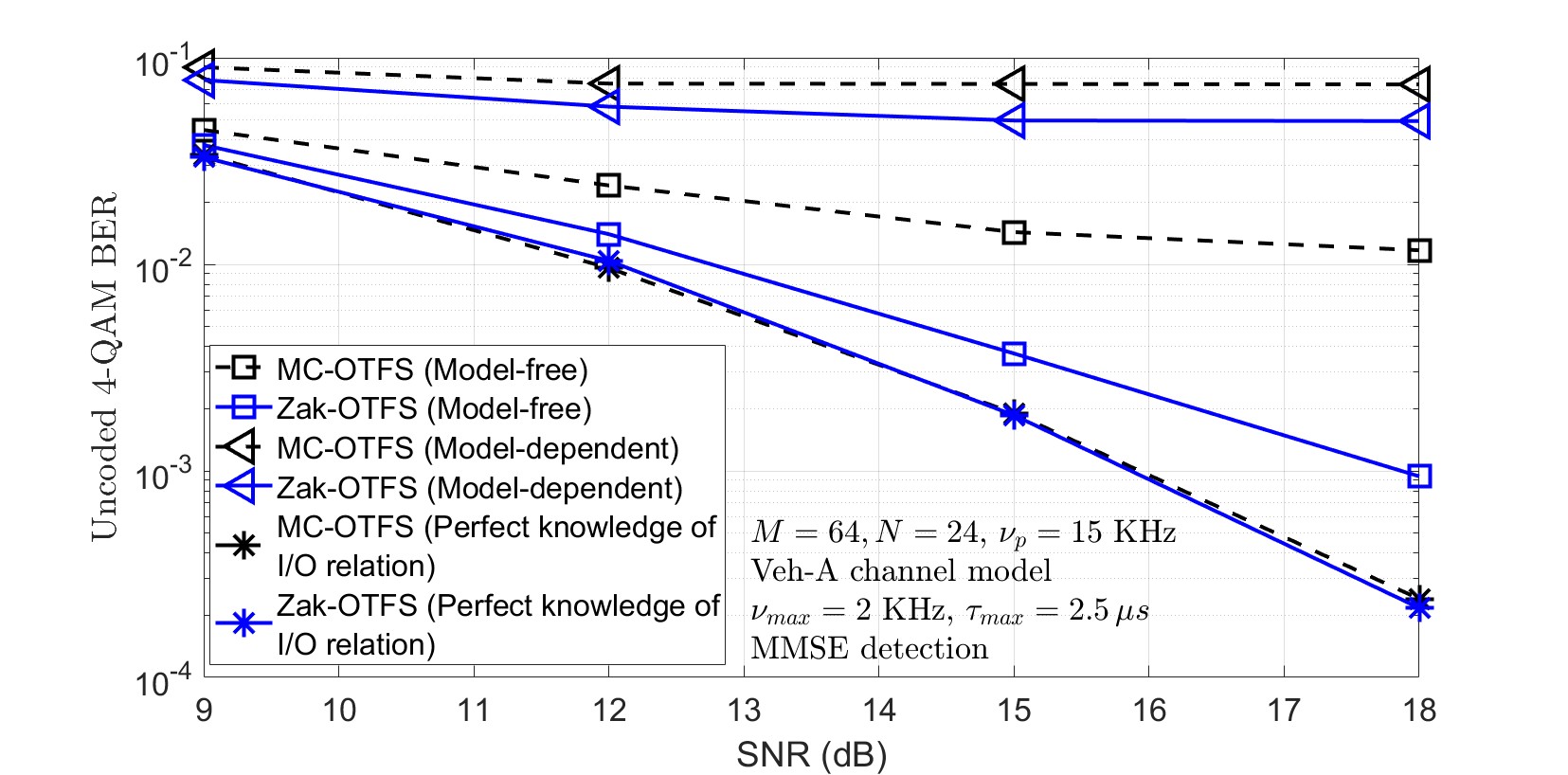}
\caption{Uncoded BER performance for Zak-OTFS and MC-OTFS on a Veh-A channel as a function of increasing SNR. The performance of model-dependent Zak-OTFS and MC-OTFS is poor since channel bandwidth and duration is not sufficient to estimate channel path gains, delays and Doppler shifts. Performance of model-free Zak-OTFS is superior to that of model-free MC-OTFS, because the Zak-OTFS I/O relation is more predictable.
}
\label{fig82}
\end{figure}

Fig.~\ref{fig82} compares BER performance of Zak-OTFS and MC-OTFS on the Veh-A channel introduced in Section \ref{secperfcsi}, and used in Fig.~\ref{fig4}. We have changed the channel parameters slightly by increasing the Doppler shift $\nu_{max}$ from $815$ Hz to $2$ KHz. For MC-OTFS, the TF window and the multicarrier signal generator are the same as those used in Fig.~\ref{fig3paper2_mcotfs} (see the above discussion). Given perfect knowledge of the I/O relation the performance of Zak-OTFS and MC-OTFS is quite similar. However, with imperfect knowledge of the I/O relation, model-free performance of Zak-OTFS is greatly superior to that of MC-OTFS because the Zak-OTFS I/O relation is more predictable than that of MC-OTFS.

\underline{Remark:} Fig.~\ref{fig82} reveals that with perfect knowledge of the I/O relation, both Zak-OTFS and MC-OTFS exhibit similar BER performance. Acquiring the I/O relation in the model-free approach amounts to estimating the channel response to a pilot at any arbitrary location. In Zak-OTFS, operating within the crystalline regime, allows to accurately predict a channel response to an arbitrary pilot from the response to a single pilot, facilitating efficient acquisition of the complete I/O relation. In contrast,
in MC-OTFS, due to the complex nature of its I/O relation, it seems that such prediction scheme is not possible, hence accurate acquisition of the complete I/O relation becomes less efficient when compared to that for Zak-OTFS. {Since accurate acquisition of the MC-OTFS I/O relation is challenging, several prior works on MC-OTFS consider the model-dependent approach where the goal is to estimate the underlying physical channel (e.g., the path gain, delay and Doppler shifts in case of a discrete multi-path channel). In comparison, for Zak-OTFS we note that it is the simplicity of its I/O relation which makes model-free approach practical.}
\begin{figure}[h]
\hspace{-7mm}
\includegraphics[width=9.9cm, height=6.5cm]{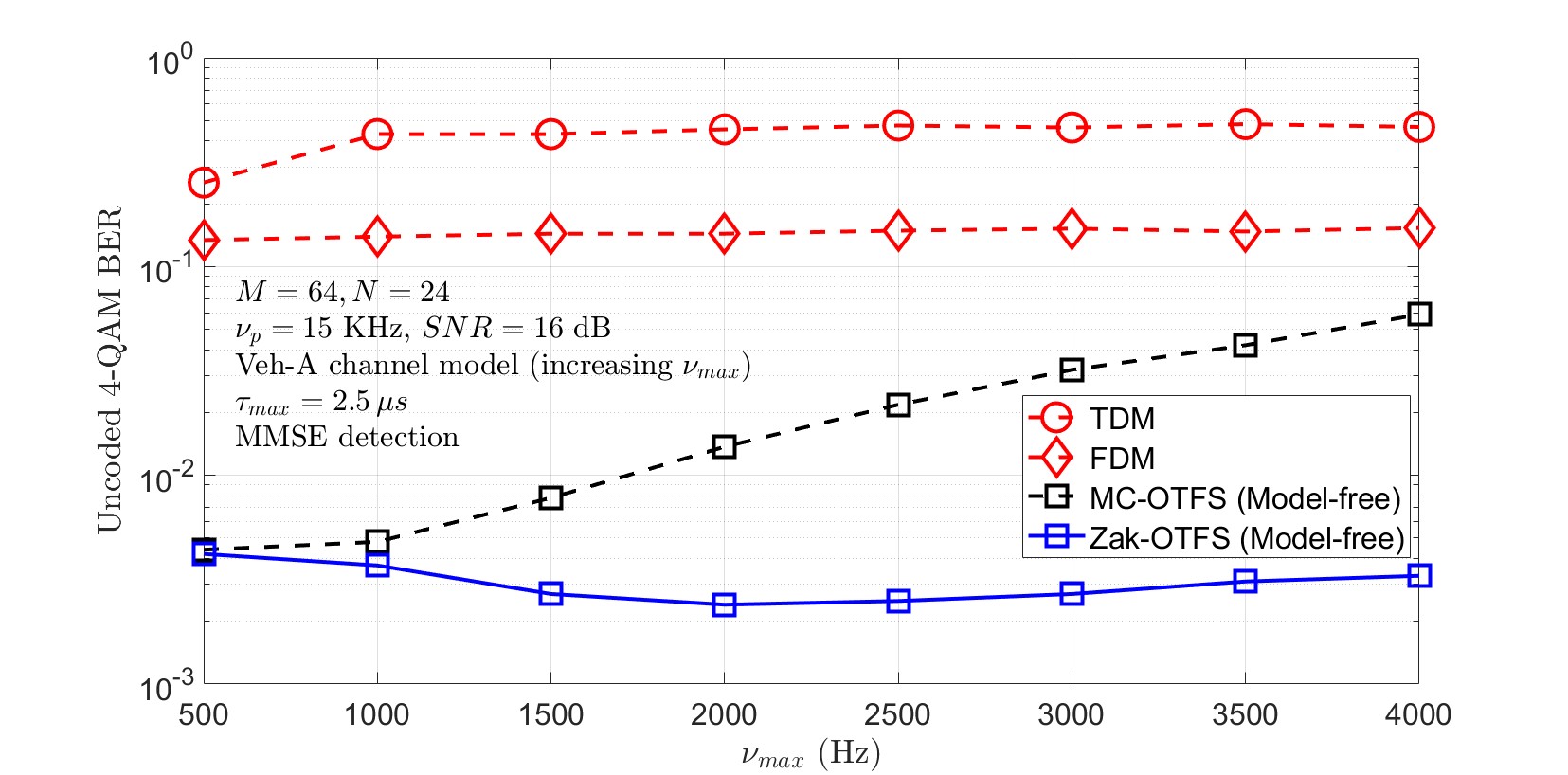}
\caption{
Uncoded BER performance of model-free Zak-OTFS and model-free MC-OTFS as a function of $\nu_{max}$ for the Veh-A channel introduced in Section \ref{secperfcsi} ($\nu_p = 15$ KHz, $M =64, N= 24$). Zak-OTFS is more robust to high Doppler spreads.
}
\label{fig15}
\end{figure}

Fig.~\ref{fig15} compares model-free performance of various modulation schemes as a function of increasing Doppler spread while operating at a fixed SNR of $16$ dB. The simulations use the Veh-A channel introduced in Section \ref{secperfcsi}. In particular, the figure illustrates that when $\nu_{max}$ is less than $1$ KHz (Doppler spread less than $2$ KHz), the BER performance of Zak-OTFS and MC-OTFS is essentially the same. However, as the Doppler spread increases, BER performance of MC-OTFS degrades steadily, while that of Zak-OTFS remains almost constant. The reason for that gap is because when the Doppler spread is high, the Zak-OTFS I/O relation is more predictable than that of MC-OTFS.

\section{The Radar Ambuguity Function}
\label{radar}
When we use a waveform to illuminate a radar scene, and we correlate the return with the transmitted waveform, the radar ambiguity function expresses the blurriness of the scene \cite{AT1985}. We begin by considering a single target and no reflectors, so that the received signal is given by
\begin{eqnarray}
\label{paper2_eqn1876}
r_{\mbox{\footnotesize{td}}}(t) & = & h \, s_{\mbox{\footnotesize{td}}}(t - \tau) \, e^{j 2 \pi \nu ( t - \tau)} \, + \, n_{\mbox{\footnotesize{td}}}(t),
\end{eqnarray}
where $s_{\mbox{\footnotesize{td}}}(t)$ is the transmitted radar waveform and $n_{\mbox{\footnotesize{td}}}(t)$
is the AWGN at the radar receiver.
{In (\ref{paper2_eqn1876}), $h$ denotes the complex channel gain suffered by the transmitted radar signal when reflected from the target and subsequently received at the radar.}
We obtain the maximum likelihood (ML) estimate for the delay $\tau$ and Doppler
$\nu$ using matched filtering at the
radar receiver. Thus $({\widehat \tau}, {\widehat \nu}) = \arg \max_{\tau, \nu} \left\vert  A_{r,s}(\tau, \nu)  \right\vert$, where
\begin{eqnarray}
 A_{r,s}(\tau, \nu) & \Define & 
\int r_{\mbox{\footnotesize{td}}}(t) \, s_{\mbox{\footnotesize{td}}}^*(t - \tau) \, e^{-j 2 \pi \nu ( t - \tau)}  dt
\end{eqnarray}
is the cross-ambiguity function between the received and the transmitted waveform. The cross-ambiguity function is also relevant for the general radar scene where there are multiple targets and reflectors. Using (\ref{paper2_eqn1}) in the expression for $A_{r,s}(\tau, \nu)$ we get
\begin{eqnarray}
\label{paper2_eqn2001}
A_{r,s}(\tau,\nu) & = &   h_{_{\mbox{\scriptsize{phy}}}}(\tau,\nu)  *_{\sigma} A_{s,s}(\tau,\nu) \nonumber \\
& &   \hspace{-3mm}  + \, \int n_{\mbox{\footnotesize{td}}}(t) \, s_{\mbox{\footnotesize{td}}}^*(t - \tau) \, e^{-j 2 \pi \nu ( t - \tau)}  dt,
\end{eqnarray}
where $*_{\sigma}$ denotes twisted convolution and $h_{_{\mbox{\scriptsize{phy}}}}(\tau,\nu)$ is the DD spreading function of the channel between the radar transmitter and receiver.
In (\ref{paper2_eqn2001}),
$A_{s,s}(\tau,\nu)$ is the
(auto-) ambiguity function of the transmitted radar waveform $s_{\mbox{\footnotesize{td}}}(t)$ and is given by
\begin{eqnarray}
\label{paper2_selfambiguityfunction}
A_{s,s}(\tau,\nu) & \Define & \int s_{\mbox{\footnotesize{td}}}(t) \, s_{\mbox{\footnotesize{td}}}^*(t - \tau) \, e^{-j 2 \pi \nu ( t - \tau)}  dt.
\end{eqnarray}
The ambiguity function $A_{s,s}(\tau, \nu)$ places fundamental limits on the \emph{blur} which constrains our ability to estimate target distance (delay) and velocity (Doppler). 
Moyal’s identity \cite{WMoran2001} captures the fundamental limits on blur by using the energy in the signal to provide a lower bound on the volume under the squared ambiguity surface.
\begin{eqnarray}
\label{paper2_Moyalidentity}
\iint \left\vert  A_{s,s}(\tau, \nu) \right\vert^2 \, d\tau d\nu & = &  \left(  \int  \vert  s_{\mbox{\footnotesize{td}}}(t) \vert^2 \, dt \right)^2.
\end{eqnarray}
Intuitively, the radar engineer aims to manipulate the ambiguity surface so that
blur is concentrated in those regions that matter least for the operational task of the radar.  

Next, we illustrate how the spread
of the ambiguity function limits the
resolution of the radar by considering a radar scene with two targets. Thus 
\begin{eqnarray}
\label{paper2_htnexample}
 h_{_{\mbox{\scriptsize{phy}}}}(\tau,\nu) & = & \sum\limits_{i=1}^2 h_i \delta(\tau - \tau_i) \delta(\nu - \nu_i),
\end{eqnarray}
where $(\tau_1, \nu_1)$, $(\tau_2, \nu_2)$ are the
delay-Doppler domain locations of the two targets. The noise-free cross-ambiguity is given by

{\vspace{-4mm}
\small
\begin{eqnarray*}
h_{_{\mbox{\scriptsize{phy}}}}(\tau,\nu) *_{\sigma}  A_{s,s}(\tau,\nu) & \hspace{-3mm} = &  \hspace{-3mm} 
h_1 A_{s,s}(\tau - \tau_1, \nu - \nu_1) e^{j 2 \pi \nu_1(\tau - \tau_1)}  \nonumber \\
& & + \, \,   h_2 A_{s,s}(\tau - \tau_2, \nu - \nu_2) e^{j 2 \pi \nu_2(\tau - \tau_2)}
\end{eqnarray*}\normalsize}
and we resolve the two targets by limiting the overlap between the two terms appearing on the R.H.S. In other words, we require the delay domain spread of $A_{s,s}(\tau, \nu)$ to be less than $\vert \tau_1 - \tau_2 \vert$ and the Doppler domain spread of $A_{s,s}(\tau, \nu)$ to be less than $\vert \nu_1 - \nu_2 \vert$.

\subsection{Ambiguity functions for TDM and FDM waveforms}
We consider a TDM pulse $s(t) = s_{\mbox{\footnotesize{td}}}(t) = \sqrt{B} \, sinc(B t)$ with bandwidth $B$, that is localized  
around $t=0$. It follows from 
(\ref{paper2_selfambiguityfunction}) that the ambiguity function in this case is given by

{\vspace{-4mm}
\small
\begin{eqnarray}
\label{paper2_eqn171}
A_{s,s}^{\mbox{\tiny{tdm}}}(\tau, \nu) &  \hspace{-3mm} = \begin{cases}
\left( 1 - \frac{\vert \nu \vert}{B} \right) \, e^{j \pi \nu \tau} \, sinc((B - \vert \nu \vert) \tau) &, \vert \nu \vert < B \\
0 &, \vert \nu  \vert \geq B \\
\end{cases}.
\end{eqnarray}\normalsize}
For a fixed $\nu$, we consider the term $sinc((B - \vert \nu \vert) \tau)$ as a function of $\tau$ and conclude that the delay spread of the TDM ambiguity function is about $1/B$. Hence, it is possible to separate two targets if their delays differ by more than $1/B$. On the other hand, the Doppler spread of the TDM ambiguity function is the bandwidth $B$. Hence, it is not possible to separate two targets with delays differing by less than $1/B$ unless the two Doppler shifts differ by more than $B$, which is unlikely in most scenarios. The TDM waveform is localized in the TD, but not in the FD, and this is the reason it is unable to separate targets in the Doppler domain.

Next, we consider an FDM pulse $s(f) = s_{\mbox{\footnotesize{fd}}}(f) = \sqrt{T} \, sinc(f T)$ with duration $T$, that is localized around $f=0$. It follows from (\ref{paper2_selfambiguityfunction}) that the ambiguity function in this case is given by

{\vspace{-4mm}
\small
\begin{eqnarray}
\label{paper2_eqnfdmradar}
A_{s,s}^{\mbox{\tiny{fdm}}}(\tau, \nu) &  \hspace{-3mm} = \begin{cases}
\left( 1 - \frac{\vert \tau \vert}{T} \right) \, e^{j \pi \nu \tau} \, sinc((T - \vert \tau \vert) \nu) &, \vert \tau \vert < T \\
0 &, \vert \tau  \vert \geq T \\
\end{cases}.
\end{eqnarray}\normalsize}
Now, the Doppler spread is small (about $1/T$) and the delay spread is large (the duration $T$), hence it is not possible to separate two targets with Dopplers differing by less than $1/T$ unless the two delay shifts differ by more than $T$. This is unlikely in most scenarios. The FDM waveform is localized in the FD, but not in the TD, and this is the reason it is unable to separate targets in the delay domain.
\begin{figure}[h]
\hspace{-4mm}
\includegraphics[width=9.6cm, height=6.4cm]{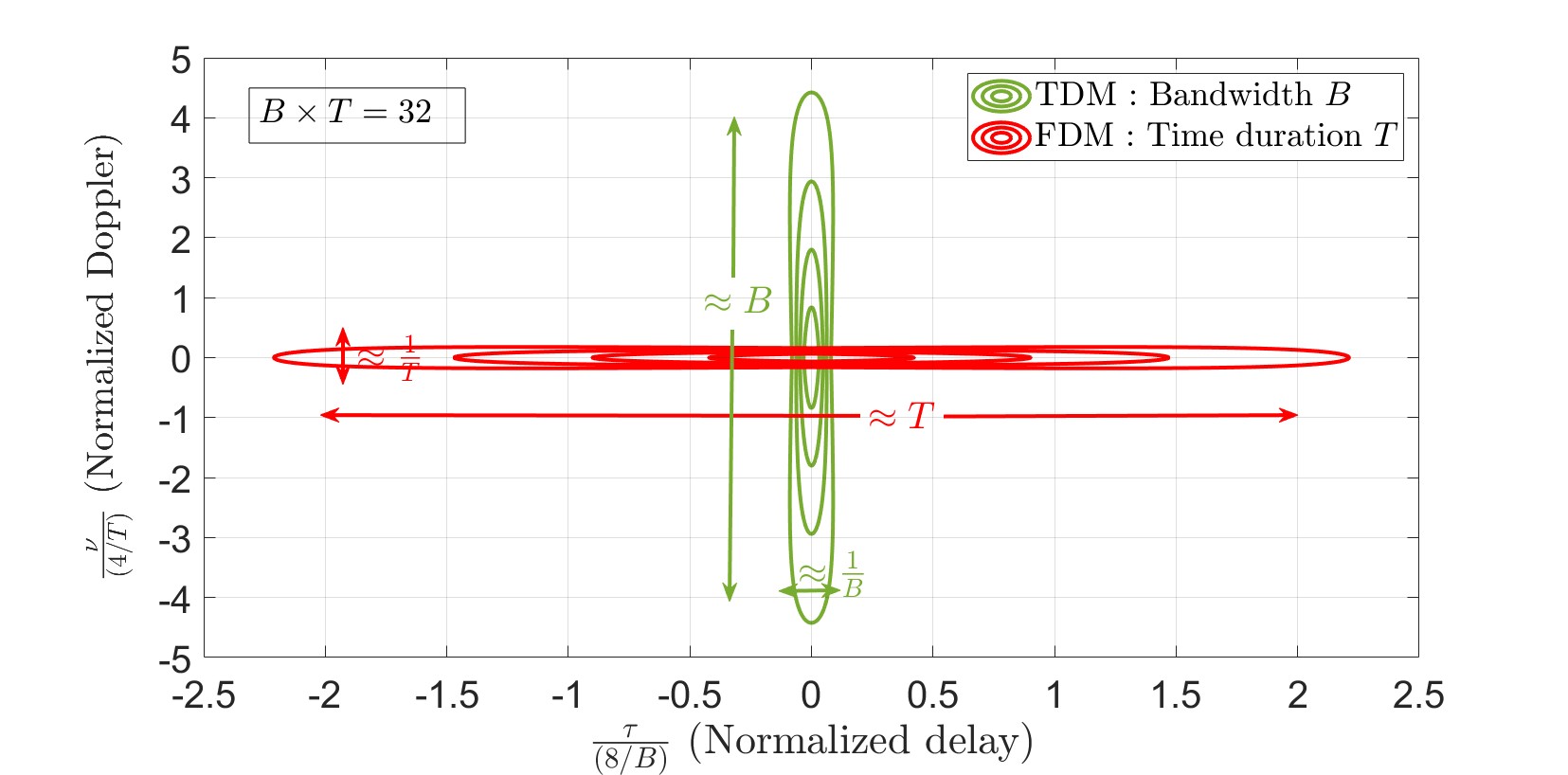}
\caption{Squared magnitude
$\left\vert A_{s,s}(\tau, \nu) \right\vert^2$ of the ambiguity functions for TDM and FDM. The TDM carrier waveform is not able to separate targets in Doppler, and the FDM carrier waveform is not able to separate targets in delay.
}
\label{fig20}
\end{figure}

\begin{figure}[h]
\hspace{-1mm}
\includegraphics[width=8.8cm, height=7.7cm]{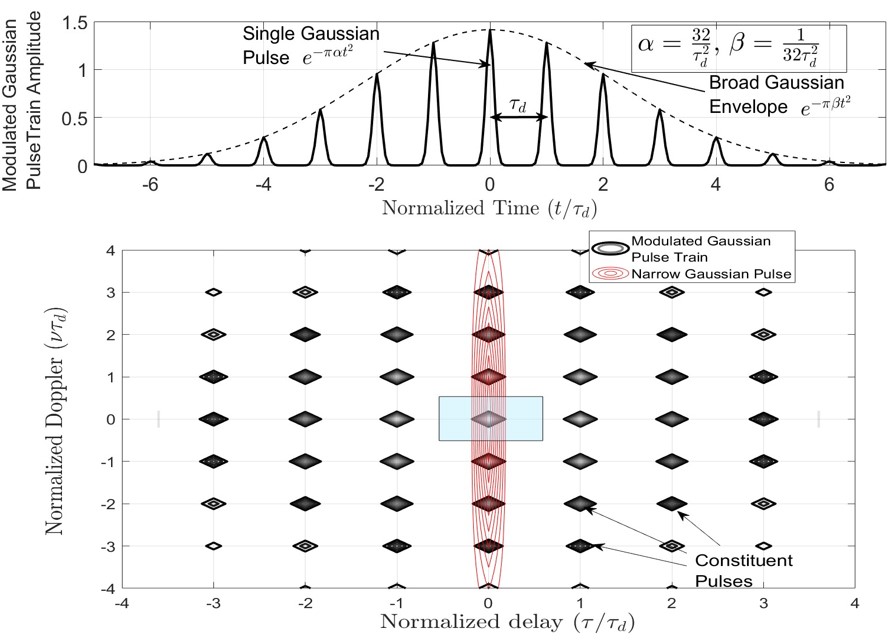}
\caption{Redistributing the squared magnitude $\vert A_{s,s}(\tau, \nu) \vert^2$ of the ambiguity function of a narrow Gaussian pulse. Modulating a train of narrow TD Gaussian pulses with a broad Gaussian envelope, produces an ambiguity function that is better able to separate targets in delay and Doppler.
}
\label{fig21}
\end{figure}

The volume under the (squared) magnitude ambiguity surface is fixed by Moyal’s identity (\ref{paper2_Moyalidentity}), but it can be redistributed to enable resolution of radar targets. This possibility was known to P. M. Woodward more than $70$ years ago. In his $1953$ book \cite{PMWoodward}, he described how the ambiguity function of a narrow Gaussian pulse (the red shaded ellipse in Fig.~\ref{fig21}) can be redistributed into several DD domain functions/pulses (shown as black ellipses in Fig.~\ref{fig21}). The trick is to modulate a train of narrow TD Gaussian pulses with a broad Gaussian envelope. There is a striking resemblance between Woodward’s waveform and the Zak-OTFS carrier waveform (pulsone), which is a train of narrow pulses modulated by a sinusoid.  

\subsection{Ambiguity function of Zak-OTFS pulsone}
The Zak-OTFS carrier waveform is a pulsone in time which realizes a quasi-periodic pulse in the DD domain at some location $(\tau_0, \nu_0)$ in the fundamental period. Recall from Part I, Table III that after filtering at the transmitter, the DD domain signal is given by

{\vspace{-4mm}
\small
\begin{eqnarray}
x_{_{\mbox{\footnotesize{dd}}}}^{w_{tx}}(\tau, \nu) & \hspace{-3mm} = & \hspace{-3mm}  w_{tx}(\tau, \nu) \, *_{\sigma} \, \nonumber \\
& & \hspace{-6mm} \sum\limits_{n,m \in {\mathbb Z}}  \hspace{-1.5mm}  e^{j 2 \pi n \nu \tau_p } \delta(\tau - n \tau_p - \tau_0) \delta (\nu - m \nu_p - \nu_0).
\end{eqnarray}
\normalsize}
Converting from the DD domain to the time domain by applying the inverse Zak transform, yields
\begin{eqnarray}
 s_{\mbox{\footnotesize{td}}}(t) & = &  \sqrt{\tau_p} \, \int_{0}^{\nu_p} x_{_{\mbox{\footnotesize{dd}}}}^{w_{tx}}(t, \nu) \, d\nu.
\end{eqnarray}
We assume the number of Doppler bins ($N$) is even, and we apply a sinc shaping filter $w_{tx}(\tau, \nu) = \sqrt{B T } sinc(B \tau) sinc(T \nu)$ to an impulse located at the origin i.e., $(\tau_0, \nu_0) = (0,0)$. The corresponding TD pulsone is given by
\begin{eqnarray}
\label{paper2_stdpulsone}
s_{\mbox{\footnotesize{td}}}(t) &  \hspace{-3.5mm} = & \hspace{-3.5mm} \sqrt{B/N} \sum\limits_{n = -\frac{N}{2}}^{\frac{N}{2} - 1}  \hspace{-2mm} sinc\left( B ( t - n \tau_p - \tau_0) \right) \, e^{j 2 \pi n \nu_0 \tau_p}.
\end{eqnarray}
\begin{figure*}
\vspace{-8mm}
\begin{eqnarray}
\label{paper2_eqnaotfs}
A_{s,s}^{\mbox{\tiny{otfs}}}(\tau, \nu) & \hspace{-3mm}  = & \hspace{-3mm} \begin{cases}
\,\, \frac{\left( 1 - \frac{\vert \nu \vert}{B}\right)}{N}  \hspace{-2mm} \sum\limits_{n_1 = -\frac{N}{2}}^{\frac{N}{2} -1} \sum\limits_{n_2 = -\frac{N}{2}}^{\frac{N}{2} -1} {\Big [} e^{j \pi \nu (\tau - (n_1 + n_2)\tau_p)} \, sinc\left( (B - \vert \nu \vert) (\tau + (n_1 - n_2) \tau_p)  \right) {\Big ]} \, &, \,\, \vert \nu \vert < B \\
\,\, 0 \, &, \,\, \vert \nu \vert \geq B
\end{cases}.
\end{eqnarray}
\end{figure*}
The corresponding ambiguity function is given in (\ref{paper2_eqnaotfs}) (see top of next page), and is illustrated in Fig.~\ref{fig22} for the same time and bandwidth constraints as the TDM and FDM waveforms illustrated in Fig.~\ref{fig20}. 
\begin{figure}
\vspace{-3mm}
\includegraphics[width=9.3cm, height=6.5cm]{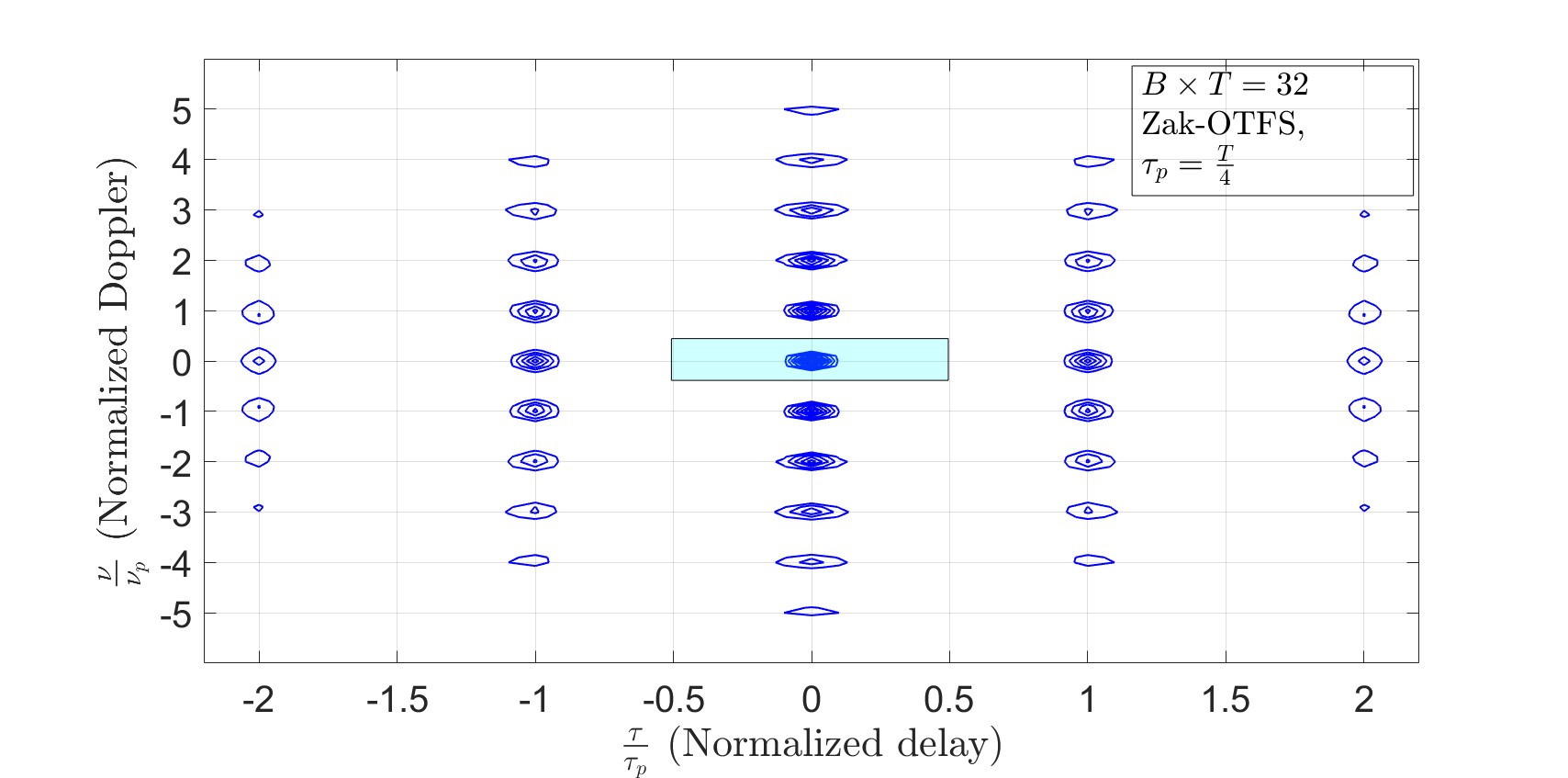}
\caption{Plot of the ambiguity function (squared magnitude)
for the Zak-OTFS carrier waveform. Simultaneous delay and Doppler resolvability can be achieved. Unambiguous target estimation is achievable in the crystalline regime. 
}
\label{fig22}
\end{figure}

The ambiguity function of the TD pulsone consists of narrow DD domain impulses separated by $\tau_p$ along the delay axis and $\nu_p$ along the Doppler axis. Each impulse has a spread of $1/B$ along the delay axis, and a spread of $1/T$ along the Doppler axis. Recall from (\ref{paper2_eqn2001}), that in the absence of noise, the cross-ambiguity function is the twisted convolution of the delay-Doppler spreading function $h_{_{\mbox{\scriptsize{phy}}}}(\tau,\nu)$ and the ambiguity function $A_{s,s}(\tau, \nu)$. We can therefore guarantee unambiguous delay–Doppler estimation by choosing $\tau_p$ to be greater than the delay spread, and $\nu_p$ to be greater than the Doppler spread. These are exactly the crystallization conditions guaranteeing predictability of the I/O relation in Zak-OTFS communication. Note that we can improve resolvability by increasing bandwidth $B$ and duration $T$.

Given an arbitrary pulse shaping filter, it follows from (\ref{paper2_selfambiguityfunction}) that the ambiguity function of the corresponding TD pulsone consists of pulses, where the pulse within the fundamental period (depicted in Fig.~\ref{fig22} as a light blue rectangle) is simply the twisted convolution of $w_{tx}(\tau, \nu)$ with its complex conjugate. In the above example, the pulse shaping filter is a sinc pulse, the Zak-OTFS pulsone is a train of sinc pulses, and the ambiguity function in Fig.~\ref{fig22} also consists of sinc pulses. In general, this structure enables precise design of ambiguity functions. 

\section{Conclusions}
\label{conc}

{
In this paper we have designed a parametric family of pulsone waveforms that can be matched to the delay and Doppler spreads of different propagation environments. We explained that a pulsone is a signal on the time domain which realizes a quasi-periodic localized function on the DD domain. The prototypical structure of a pulsone is a train of pulses  modulated by a tone.
We have put special emphasis on system analysis in the situation when the pulsone parameters matches the environment channel parameters, in the sense that, the delay period of the pulsone is greater than the delay spread of the channel, and the Doppler period of the pulsone is greater than the Doppler spread of the channel. We refer to this condition as the crystallization condition.

We have explained that pulsones constitute a universal parametric family of waveforms which includes as special cases conventional TDM and FDM waveform. Specifically, a TDM waveform is a pulsone with infinite delay period and zero Doppler period. Such a waveform perfectly matches to a delay-only propagation environments. Reciprocally, an FDM waveform is a pulsone with infinite Doppler period and zero delay period. Such a waveform perfectly matches to Doppler-only propagation environments.

On the negative side, we have explained, in the context of communication, how and why the I/O relation of sampled system becomes non-predictable and fading when the parameters of the pulsone do not match the propagation environment, i.e., they do not satisfy the crystallization condition. Specifically, we have shown that the phenomena of non-predictability results from aliasing in the DD domain, which, in turn, occurs when one of the channel spreads are greater than the corresponding pulsone period. In the context of radar sensing, we have shown that ambiguity - that is, inability to separate resolvable channel reflections, occurs when one of the propagation spreads is greater than the corresponding pulsone period - again, violation of the crystallization condition.

On the positive side, we have explained how and why a sampled communication system yields superior performance when operated in the crystalline regime, i.e,  when the delay spreads of the channel are considerably smaller than the periods of the pulsone. Specifically, we have explained that operation in the crystalline regime provides two mechanisms for improved performance: one mechanism minimizes DD domain aliasing through proper choice of the periods and shaping filters, and a second mechanism maximizes diversity exploitation by resolving reflections through suitable choice of shaping filters. In the context of radar sensing, minimizing DD domain aliasing translates to reducing ambiguity among resolvable reflections, and maximizing resolvability translates to increased resolution. 

Another important implication, in the context of communication, of operation with pulsones in the crystalline regime is that, due to its predictability, the I/O relation of a sampled communication system can be learnt directly without the need to know the parameters of the underlying channel. This opens up the possibility of a model-free mode of operation, which is especially useful when channel estimation is out of reach.

Another contribution of this paper is a detailed comparison between Zak-OTFS and its multicarrier approximation, which we refer to as MC-OTFS, that has been the focus of almost all research attention so far. We compared the two modulation schemes both on theoretical grounds and on performance grounds. On the theoretical side, we have shown that the I/O relation of MC-OTFS is less predictable than that of Zak-OTFS, which implies that as the Doppler spread increases, the BER performance of MC-OTFS is inferior to that of Zak-OTFS. This suggests that MC-OTFS is less adapted to model-free mode of operation than Zak-OTFS.

In conclusion, convergence of communications and sensing in 6G and beyond has focused research attention on the design of carrier waveforms that support both applications. Noticing that an environment, be it a radar scene or a communication medium is characterized by its delay and Doppler spreads. The underlying message of this paper is that in both contexts - radar and communication, it is beneficial to choose the periods of the pulsone to be greater than the spreads of the environment, i.e., one should operate in the crystalline regime. 

}

\section{Acknowledgements}
The authors would like to thank the Associate Editor, Suhas Diggavi
and the anonymous reviewers for their very helpful suggestions.
The work of Saif Khan Mohammed was supported by the Prof. Kishan and Pramila Gupta Chair at I.I.T. Delhi. A. Chockalingam acknowledges the support from the J. C. Bose National Fellowship, Science and Engineering Research Board, Department of Science and Technology, Government of India. The work of Robert Calderbank is supported in part by the Air Force
Office of Scientific Research under Grants FA 87520-20-2-0504 and FA
9550-20-1-0266, and by the National Science Foundation under Grant
FAIN-2148212.

\begin{IEEEbiography}[{\includegraphics[width=1.0in,height=1.25in]{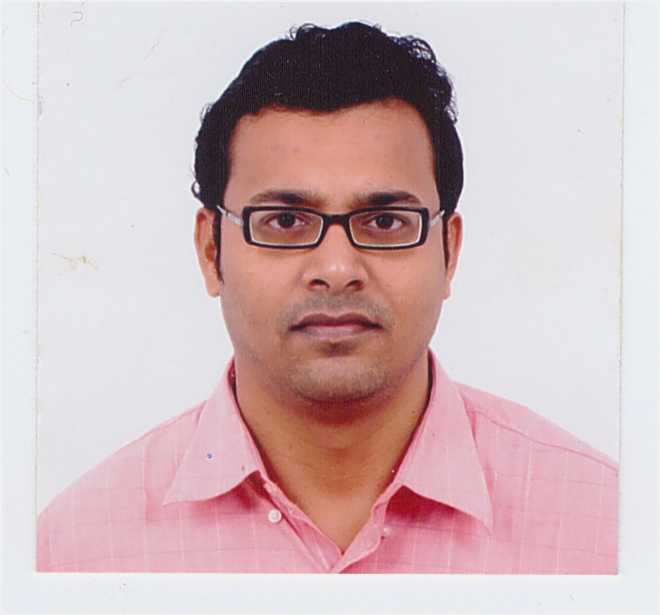}}]{Saif Khan Mohammed}
is a Professor with the
Department of Electrical Engineering, Indian Institute of Technology Delhi (IIT Delhi). He is also associated with the Bharti School of Telecommunication Technology and Management, IIT Delhi. He currently holds the Prof. Kishan and Pramila Gupta Chair at IIT Delhi.
He received the B.Tech. degree in Computer Science and Engineering from IIT Delhi,
New Delhi, India, in 1998, and the Ph.D. degree from the Electrical and Communication
Engineering Department, Indian Institute of Science, Bangalore, India, in 2010.
From 2010 to 2011, he was a
Post-Doctoral Researcher at the Communication Systems Division (Commsys), Electrical Engineering
Department (ISY), Linkoping University, Sweden. He was an Assistant Professor at Commsys, from
September 2011 to February 2013. His main research interests include, waveforms for high
mobility scenarios in sixth generation (6G) communication systems, wireless communication
using large antenna arrays, coding and signal processing for wireless communication
systems, information theory, and statistical signal processing. He currently serves as an
Editor for IEEE Transactions on Wireless Communications and in the past he has served as an Editor for IEEE
Wireless Communications Letters and Physical Communication journal (Elsevier). He holds four
granted U.S. patents in multi-user detection and precoding for multiple-input multiple-output
(MIMO) communication systems. He received the 2017 NASI Scopus Young Scientist Award and the
Teaching Excellence Award at IIT Delhi for the year 2016–2017. He was also a recipient of the
Visvesvaraya Young Faculty Fellowship from the Ministry of Electronics and IT, Government of India,
from 2016 to 2019. Contact him at saifkmohammed@gmail.com.
\end{IEEEbiography}

\begin{IEEEbiography}[{\includegraphics[width=1in,height=1.25in]{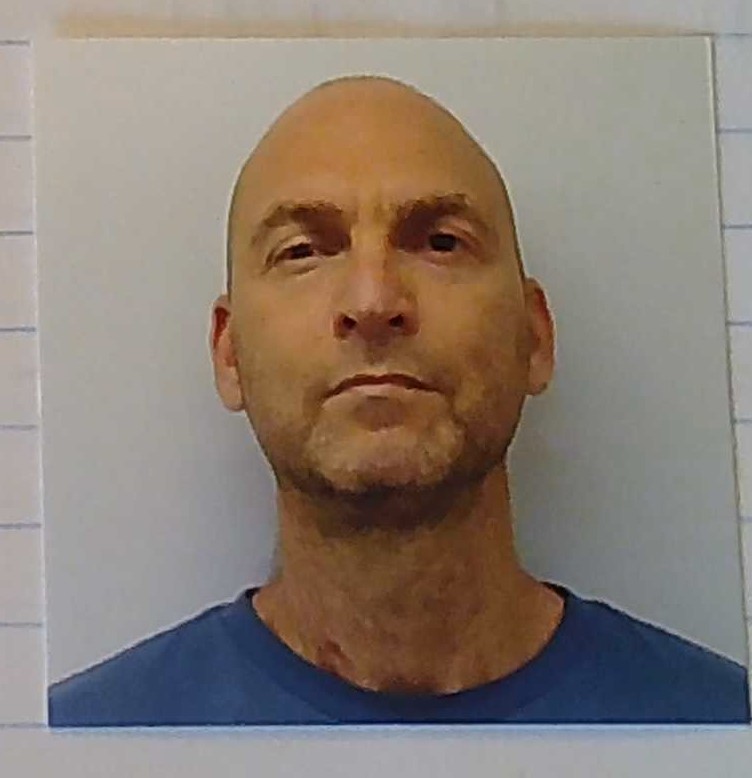}}]{Ronny Hadani}
is an associate professor in the Mathematics Department of the University of Texas at Austin. He also serves as the Chief Science Officer at Cohere Technologies. He holds a PhD in pure mathematics from Tel-Aviv University and a Master degree in applied mathematics from The Weizmann Institute of Science. Contact him at hadani@math.utexas.edu. 
\end{IEEEbiography}

\begin{IEEEbiography}[{\includegraphics[width=1in,height=1.25in]{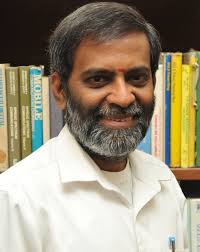}}]{Ananthanarayanan Chockalingam}
was born in Rajapalayam, Tamil Nadu, India. He received the B.E. (Honors) degree in electronics and communication engineering from the P. S. G. College of Technology, Coimbatore, India, in 1984, the M. Tech. degree in electronics and electrical communications engineering (with specialization in satellite communications) from the Indian Institute of Technology, Kharagpur, India, in 1985, and the Ph.D. degree in electrical communication engineering (ECE) from the Indian Institute of Science (IISc), Bangalore, India, in 1993. During 1986 to 1993, he worked with the transmission R\&D division of the Indian Telephone Industries Limited, Bangalore. From December 1993 to May 1996, he was a Postdoctoral Fellow and an Assistant Project Scientist in the department of electrical and computer engineering, University of California, San Diego. From May 1996 to December 1998, he served Qualcomm, Inc., San Diego, CA, as a Staff Engineer/Manager in the systems engineering group. In December 1998, he joined the faculty of the department of ECE, IISc, Bangalore, India, where he is a Professor, working in the area of wireless communications.

Dr. Chockalingam is a recipient of the Swarnajayanti Fellowship from the Department of Science and Technology, Government of India. He served as an Associate Editor of the IEEE TRANSACTIONS ON VEHICULAR TECHNOLOGY, and as an Editor of the IEEE TRANSACTIONS ON WIRELESS COMMUNICATIONS. He served as a Guest Editor for the IEEE JOURNAL ON SELECTED AREAS IN COMMUNICATIONS (Special Issue on Multiuser Detection for Advanced Communication Systems and Networks), and for the IEEE JOURNAL OF SELECTED TOPICS IN SIGNAL PROCESSING (Special Issue on Soft
Detection on Wireless Transmission). He is a Fellow of the Indian National Academy of Engineering, the National Academy of Sciences, India, the Indian National Science Academy, and the Indian Academy of Sciences. Contact him at achockal@iisc.ac.in.
\end{IEEEbiography}

\begin{IEEEbiography}[{\includegraphics[width=1in,height=1.25in]{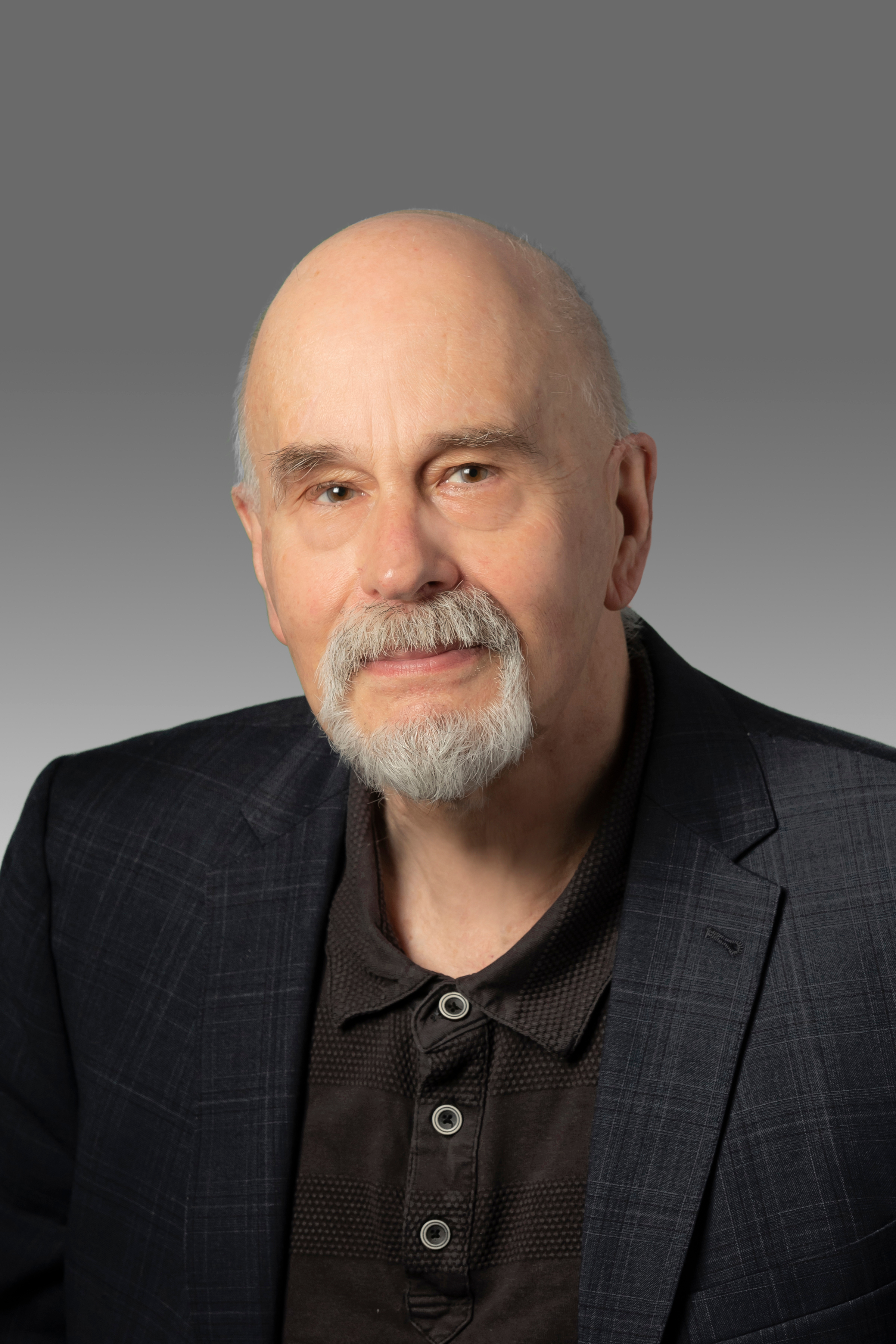}}]{Robert Calderbank}
directs the Rhodes Information initiative at Duke University, where he is a Distinguished Professor of Electrical and Computer Engineering, Computer Science and Mathematics.
 
He started his career in the Mathematical Sciences Research Center at Bell Labs, and he left AT\&T in 2003 as Vice President for Research. Dr. Calderbank directed the Program in Applied and Computational Mathematics at Princeton University before joining Duke University in 2010.  He was elected to the National Academy of Engineering in 2005, and to the American Academy of Arts and Sciences in 2022. Dr. Calderbank received the 2015 Hamming Medal, and the 2015 Shannon Award.
 
Dr. Calderbank is known for contributions to voiceband modem technology at the dawn of the internet, and for contributions to wireless communication that are incorporated in billions of cell phones. He has also made contributions to quantum error correction that provide a foundation for fault tolerant quantum computation. Contact him at robert.calderbank@duke.edu.
\end{IEEEbiography}

\end{document}